\documentclass{aa}  
\usepackage{graphicx}
\usepackage{derivative}
\usepackage{xcolor}
\usepackage{float}
\usepackage{rotating}
\usepackage{booktabs}
\usepackage{mathtools}
\usepackage{placeins}
\usepackage{longtable}
\usepackage{supertabular}
\usepackage{adjustbox}
\usepackage{supertabular}
\usepackage{multirow}
\usepackage{tabularx}
\usepackage{url}
\usepackage{esdiff}
\usepackage{amsmath}
\usepackage{siunitx}
\usepackage[switch]{lineno}
\usepackage{hyperref}
\usepackage{lscape}
\hypersetup{
  colorlinks=true,
    linkcolor=blue,
    citecolor=blue,
    urlcolor=magenta}

\newcommand{\kep}{{\it Kepler}\xspace}
\newcommand{\prot}{$P_\text{rot}$\xspace}
\newcommand{\porb}{$P_\text{orb}$\xspace}

\newcommand{\teff}{$T_\text{eff}$\xspace}
\usepackage{natbib}
\bibpunct{(}{)}{;}{a}{}{,}

\usepackage[varg]{txfonts}
\begin{document}

   \title{Non-transiting exoplanets as a means of understanding star-planet interactions in close-in systems}
    
   \titlerunning{}

   \author{C. Gourvès\inst{1,2}
           \and
           S.N.~Breton\inst{2}
           \and
           A.~Dyrek\inst{3,1}
           \and
           A.F.~Lanza\inst{2}
           \and 
           R.A.~García\inst{1}
           \and
           S.~Mathur\inst{4,5}
           \and
           {\^A}.R.G.~Santos\inst{6}
           \and
           A.~Strugarek\inst{1}
          }
    \institute{Universit\'e Paris-Saclay, Universit\'e Paris Cit\'e, CEA, CNRS, AIM, 91191, Gif-sur-Yvette, France \email{clemence.gourves@cea.fr}
    \and
    INAF – Osservatorio Astrofisico di Catania, Via S. Sofia, 78, 95123 Catania, Italy
    \and
    Space Telescope Science Institute, 3700 San Martin Drive, Baltimore, MD 21218, USA
    \and
    Instituto de Astrof\'{\i}sica de Canarias, La Laguna, Tenerife, Spain
    \and 
    Departamento de Astrof\'{\i}sica, Universidad de La Laguna, La Laguna, Tenerife, Spain
    \and
    Instituto de Astrof\'isica e Ci\^encias do Espa\c{c}o, Universidade do Porto, CAUP, Rua das Estrelas, PT4150-762 Porto, Portugal
    }

   \date{}

 \abstract{
Previous studies showed evidence of a dearth of close-in exoplanets around fast rotators, which can be explained by the combined action of intense tidal and magnetic interactions between planets and their host star. Detecting more exoplanets experiencing such interactions, with orbits evolving on short timescales, is therefore crucial to improve our understanding of the underlying physical mechanisms.
For this purpose, we performed a new search for close-in non-transiting substellar companions in the \kep data, focusing on orbital periods below 2.3 days. We focused on main-sequence solar-type stars and subgiant stars for which a surface rotation period was measured.  
For each star, we looked for an excess in the power spectral density of the light curve, which could correspond to the signature of a close-in non-transiting companion. We compared our candidates with existing catalogues to eliminate potential contaminants in our sample, and we visually inspected the phase-folded light curve and its wavelet decomposition.
We identify 88 stars exhibiting a signature consistent with the presence of a close non-transiting substellar companion. We show that the objects in our sample are located mostly within the dearth zone, emphasising the importance of performing follow-up of such systems in order to gather observational evidence of star-planet interactions.}

 
\maketitle

\section{Introduction \label{section:introduction}}

Since the first observation of a planet orbiting a solar-type star \citep{Mayor1995}, more than 5,500 exoplanets have been confirmed, the majority detected by the \kep satellite with the transit method \citep{Borucki2010}. In particular, this has opened the possibility of studying star-planet interactions shaping orbital architecture both at the scale of individual star-planet systems \citep[e.g.][]{Patra2017,2024Barker} and for larger samples \citep[e.g.][]{McQuillan2013b,Messias2022,Garcia2023}.  

Observing close-in planets, which exhibit orbital periods shorter than 1 day and experience strong tidal \citep[e.g.][]{Mathis2018} and magnetic \citep[e.g.][]{Strugarek2018} interactions with their host star, is crucial to calibrate dissipation processes related to these phenomena. Unfortunately, ultra-close-in planets, depending on the planetary mass, its orbital period, and the rotational period of the star, may be pushed away or experience orbital decay and disruption on very short timescales \citep[e.g.][]{Ahuir2021}. This means that we are more likely to observe planets that are less dynamically perturbed and located further away from their host. Among all the planets detected using the transit method, approximately 97\% have orbital periods exceeding one day.\footnote{\label{footnote:nasa_database} According to the NASA Exoplanet Archive, as of April 6, 2025. The database can be accessed
at \url{https://exoplanetarchive.ipac.caltech.edu/index.html}.} This highlights the need for alternative detection methods to study ultra close-in planets. Statistically, non-transiting systems are likely far more numerous than transiting ones, suggesting that observing them could increase our chances of detecting planets very close to their star.

\begin{figure}[htbp!]
    \centering
   \includegraphics[width=0.45\textwidth]{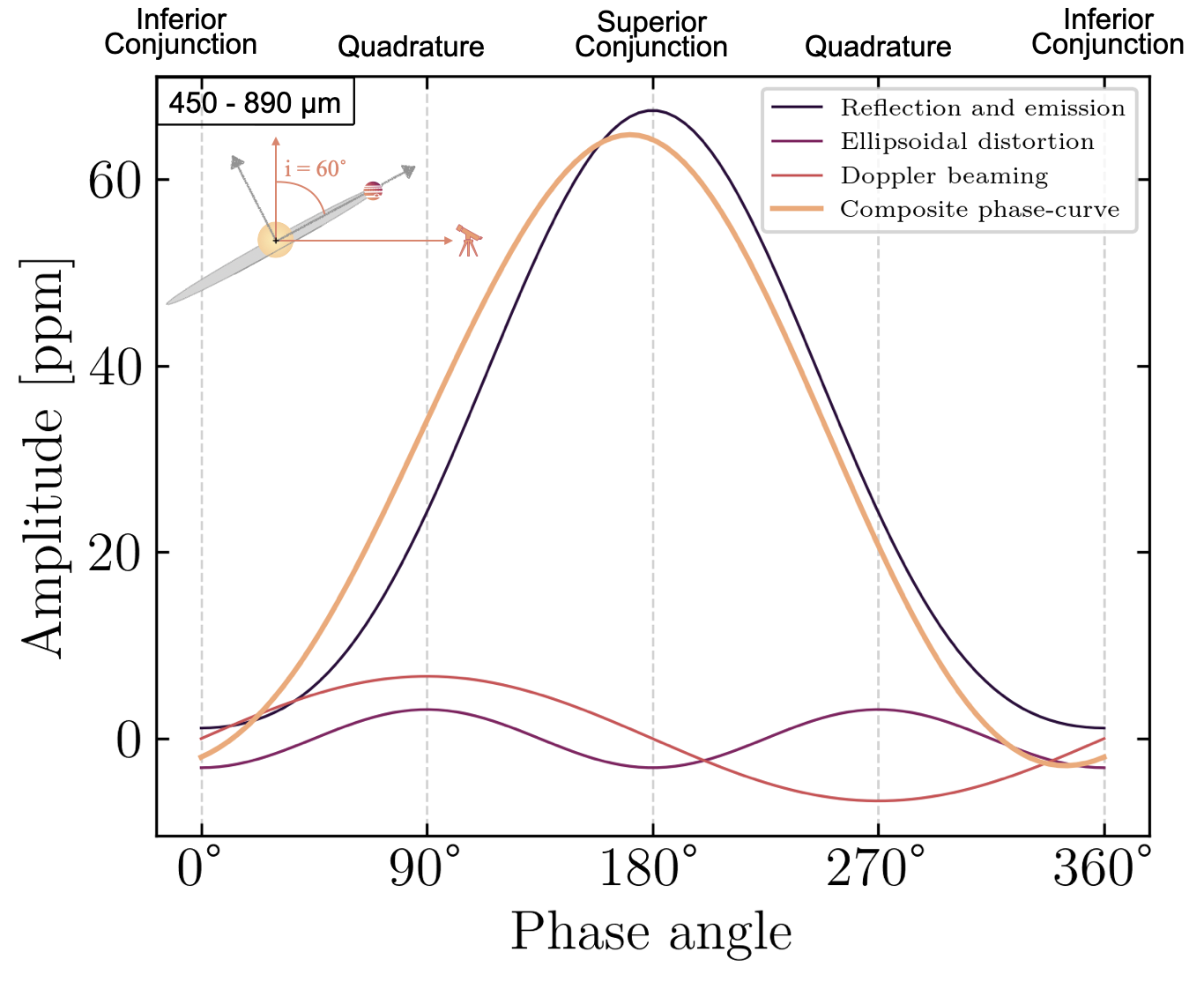}
   \caption{Simulated contributions of the three modulations to the phase-curve amplitude according to the phase angle. The modulation linked to atmospheric processes is shown in black and includes reflection and emission modulations. The ellipsoidal distortion is shown in purple, while the Doppler boosting is in orange. The overall contribution is shown by the composite phase-curve in yellow. Here, an eclipse would occur at the inferior conjunction, while the transit would happen at the superior conjunction with the observer.}
   \label{fig:composantes}
\end{figure}

Having rather large modulations in their orbital phase curve, close-in planets can be detected with photometric observations, even if they are not transiting. Until now, only a few non-transiting companions have been found \citep{Millholland2017,LilloBox2021,2024Cullen}. These modulations are attributed to three main physical processes \citep{Shporer2017}. The first is atmospheric, encompassing stellar light reflected by the companion's atmosphere and thermal emission, due to heating of the companion's atmosphere. The second is the tidal ellipsoidal distortion, induced by the tidal forces exerted by the companion on its host star. The third process is the Doppler boosting effect, caused by variations in the radial velocity between the object and the observer, leading to a modulation in the observed flux of the star.
Fig.~\ref{fig:composantes} shows an example of the three modulation components for a non-transiting planet, as well as the composite phase-curve, generated using Eq. (4), (7), (9), and (10) from \citet{Shporer2017} and a black body emission law computed at the planetary equilibrium temperature, as a function of the star-planet-observer phase angle \citep{Seager2010}. A realistic set of stellar and planetary parameters was chosen, analogous to the ones expected for a simplified bookcase-like system. A hot Jupiter with a radius of $1 \ R_{\rm jup}$, a mass of $1 \ M_{\rm jup}$, and an orbital period of 2 days was considered. The eccentricity was set to zero, while the inclination of the orbit on the line of sight was 60°, the maximum value taken for non-transiting systems \citep{Winn2014}. We opted for a Bond albedo of 0.5 \citep[e.g.][]{Seager2010}, which is typical for a hot Jupiter-type planet, and picked an atmospheric recirculation rate of 50\% \citep{Fortney2021}. We chose $1 \ \rm M_\odot$, $1 \  \rm R_\odot$, and $T_{\rm eff} = 6000$ K as stellar parameters. In this configuration, the atmospheric processes outweigh the other two phenomena.

While the atmospheric processes are mostly driven by stellar irradiation, Doppler boosting and tidal ellipsoidal distortion are induced by the bodies' gravitational interaction, and hence they depend on the stellar and planet masses. The greater the companion's orbital velocity and mass, the stronger the Doppler boosting. However, depending on the mass ratio between the star and the planet, for short orbital periods, the ellipsoidal distortion effect might outweigh the other two phenomena in amplitude. 

The layout of the paper is as follows. In Sect.~\ref{sec:data_method} we present the sample of \kep light curves we consider, and we explain how we look for a non-transiting companion signature. In Sect.~\ref{sec:fit of the photometric data}, we perform a fit of the photometric modulations to extract additional constraints on the candidates' global parameters. In Sect.~\ref{section:conclusion}, we discuss the perspectives that the characterisation of this sample may open.

\section{Observational data and search methodology \label{sec:data_method}}

In mono-band photometry and when blindly searched for in a large sample, non-transiting companion signatures may easily be confused with modulations arising from stable active longitudes of fast-rotating young stars. Therefore, prior knowledge of the rotation of the stellar host is necessary to undertake our analysis.
To avoid confusing exoplanetary signals with the rotation period \citep[e.g.][]{Garcia2014}, we considered the catalogue by \citet{Santos2019, Santos2021}, which consists of 55,232 main-sequence and subgiant FGKM stars. 

\begin{figure}[htbp!]
   \includegraphics[width=0.53\textwidth]{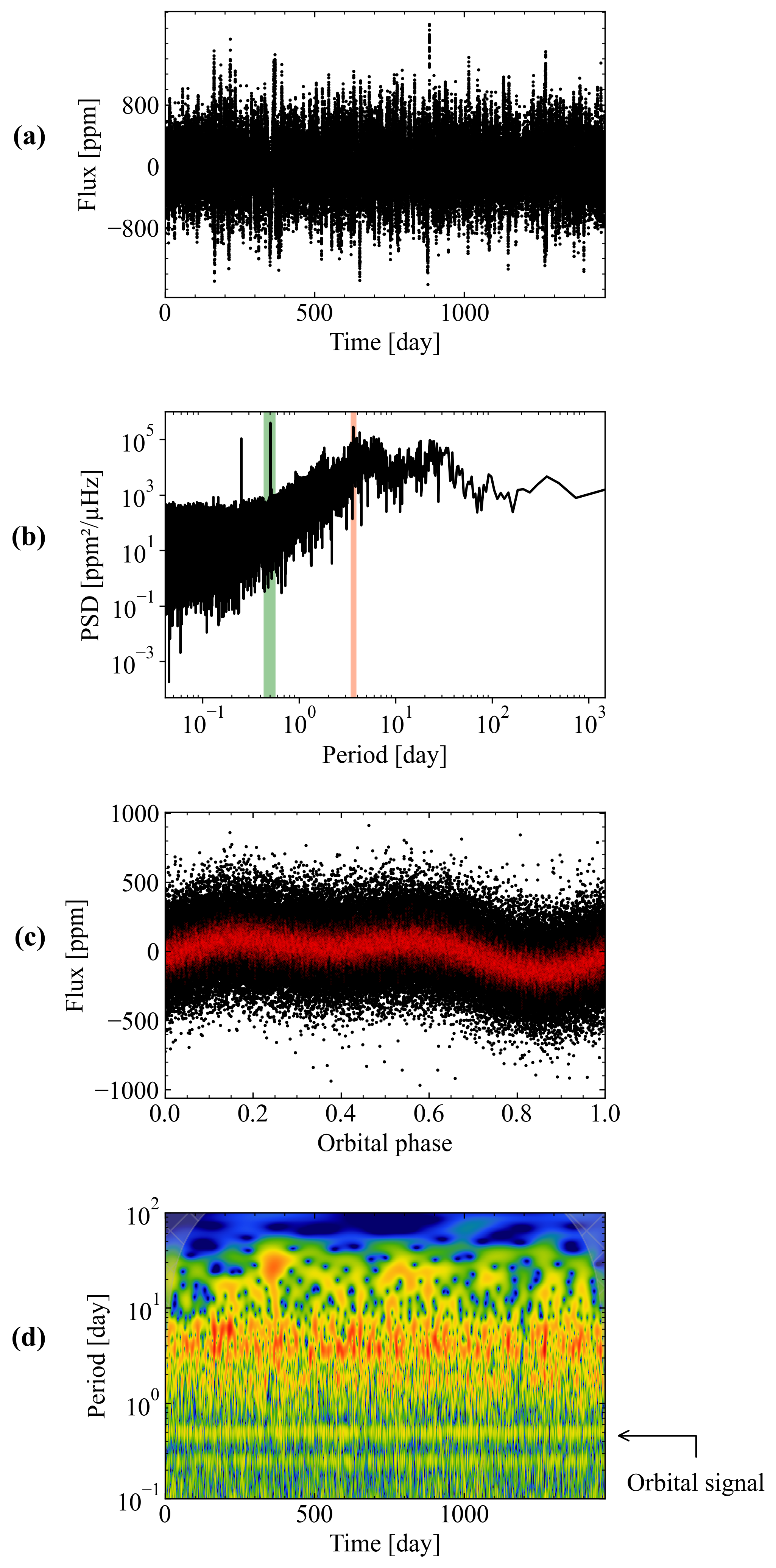}
   \caption{Flux analysis of KIC 5697777. (a) Light curve calibrated according to the KEPSEISMIC method and filtered at 20 days. (b) Power spectral density of the light curve. The rotation period of the star is taken from \citet{Santos2021} and is located within the orange area. The green area shows the high-amplitude signal peak at around 0.5 days. The second high-amplitude signal peak located at around 0.25 days is the second harmonic of the orbital signal. (c) Phase-folding of the light curve with an additional filter to remove the stellar rotational contribution. The red curve is a smoothing of the black curve, obtained with a triangular filter. (d) Wavelet power spectrum of the light curve. A high-amplitude periodicity is highlighted by orange and red, whereas yellow and blue denote a low-power amplitude. The yellow bands with a constant period throughout time correspond to the high-amplitude peak and its second harmonic, identified in panel (b). The reddish band corresponds to the rotation period and its harmonics, highlighted in orange in panel (b). The shaded area represents the cone of influence of the reliable periodicities.}
   \label{fig:KIC_example}
\end{figure}

To illustrate the procedure we describe below, we selected KIC 5697777, an F-type star with a rotation period of 3.90 days. The results obtained for this star are presented in Fig. \ref{fig:KIC_example} and are discussed in this section. 

\subsection{Unravelling signatures of non-transiting objects \label{sec:psd}}

We used \kep light curves calibrated with the KEPSEISMIC method\footnote{The light curves can be downloaded at the following address: \url{https://archive.stsci.edu/prepds/kepseismic/}.} \citep{Garcia2011, Pires2015}, whereby signals with a period longer than 20 days were filtered out \citep[see][]{Santos2019} in order to be less affected by the instrumental noise and because we are interested in close-in objects. The light curve of KIC 5697777 is presented in panel (a) of Fig. \ref{fig:KIC_example}.

The power spectral density (PSD) of a stellar light curve (Fig.~\ref{fig:KIC_example}, panel b) allows for the decomposition of the signal into its various frequency components. Identifying a high-amplitude peak, typically from a few tens to a few thousands of parts per million, is then akin to detecting a peak that corresponds to the passage of a non-transiting object. As is seen in Fig.~\ref{fig:composantes}, if tidal interactions dominate the flux modulation, the second harmonic of the orbital signal is expected to be the largest peak of the harmonic pattern (when still excluding the stellar rotation period signal). Nevertheless, it is the frequency of the first harmonic that corresponds to the orbital period of the non-transiting companion object.

To identify a non-transiting object signal, we computed the PSD of each of the 55,232 stars' light curves. To remove false positives due to low-inclination stellar binary systems, we focused on systems with a signal at a period distinct from the rotation period of the targeted star. In a close binary stellar system, tidal effects typically tend to synchronise the rotational periods of the two stars \citep[e.g.][]{1975Zahn}. In contrast, a planetary-mass companion generally lacks sufficient mass to induce such synchronisation. The aim here is to identify significant high-amplitude peaks in the PSD, different from the peak associated with the stellar rotation period. Given that only close-in objects should be detectable with our method, we focused on objects with short orbital periods and we searched for peaks with a frequency higher than 5 $\mu \textrm{Hz}$ (2.3 days). Below $5 \ \rm \mu \textrm{Hz}$, the main signal in the PSD of a main-sequence solar-like star usually comes from the surface rotational modulation, granulation, and activity. In addition, because the object would be too distant from the star, the phase curve modulation amplitude would be too low to be detectable in visible photometry \citep[e.g.][]{Shporer2017}. 
A high-amplitude peak with a frequency higher than $5 \ \mu$Hz  can then be associated with the presence of a non-transiting object. If we know the stellar rotation period and its harmonics, we do not expect any additional excess power of this nature associated with the dynamics of the star alone, when considering a main-sequence solar-type star \citep[e.g.][]{Garcia2019}.

A threshold limit was applied in the PSDs, to distinguish sharp high-amplitude peaks associated with non-transiting companions from background noise. The statistical distribution of the PSD function is a $\chi^2$ distribution with two degrees of freedom \citep{Woodard1984}. Following \citet{Appourchaux2000}, we calculated the power level at which a peak due to noise has a probability, $p_{\rm det}$, of appearing over a frequency range, $\Delta_{\rm det}$. Since the number of considered bins in the bandwidth will increase with the length of the time series, this threshold depends on the observation time, $T$, which is about 4 years for the \kep observations. The ratio of the statistical $\chi^2$ threshold to the average signal amplitude, $\Tilde{s}$, is given by $s_{\rm det}/\Tilde{s} \approx \textrm{ln}(T) + \textrm{ln}(\Delta_{\rm det}) - \textrm{ln}(p_{\rm det})$, with $p_{\textrm{det}}$ the detection probability threshold, $T$ in $\mu$s and $\Delta_{\rm det}$ in microhertz. Here, $\Delta_{\rm det}$ is within the range of 5 to $284 \ \mu \textrm{Hz}$, the Nyquist frequence of \kep observations.
After some empirical tests, we decided to consider a $s_{\rm det}/\Tilde{s}$ value of approximately 260, which corresponds to an extremely low detection probability threshold, $p_{\textrm{det}}$, of $10^{-100}$ (this corresponds to a sigma of 259). This choice allowed us to filter out most of the signal related to the harmonics of stellar rotation, while preserving the high-amplitude peaks we were looking for.

We computed $s_{\rm det}/\Tilde{s}$ by dividing the PSD by a factor of $\alpha = (9/8)^3 \times \mathrm{median} \, ({\rm PSD})$, corresponding to the corrected mean of the PSD. Nevertheless, the background signal of the PSD is non-uniform, predominantly influenced by the convective signal and the magnetic activity of the star, which results in higher amplitudes at low frequencies compared to high frequencies \citep{Garcia2019}.
Hence, adopting such a low probability threshold, $p_{\rm det}$, reduces the likelihood of obtaining false positives and improves the accuracy of our signal analysis. With this method, we preselected 4,788 stars.\\
Subsequently, we conducted a visual inspection of the PSDs displaying prominent peaks distinct from the stellar rotation period. This is illustrated in panel (b) of Fig. \ref{fig:KIC_example}, in which the PSD exhibits a high-amplitude peak around 0.52 days (green area), while the stellar rotation period of KIC 5697777 corresponds to a period of 3.9 days (orange area). For our remaining stars, the PSDs presenting peaks that are part of the harmonic chain of the rotation period are excluded from the catalogue. However, we decided to keep in our sample targets for which the detected peak lies within the harmonic pattern of surface rotation but exhibits a clearly distinct morphology with respect to it; that is, for which we observe a sharp peak amidst the wide rotation harmonics. This analysis step reduced the considered sample to 362 stars.

\subsection{Comparison with existing catalogues \label{sec:comparison with the litterature}}

To eliminate potential contaminants in our sample, we conducted a thorough cross-check with existing catalogues. The aim was to remove any stars previously identified as hosting binaries, brown dwarfs, transiting exoplanets, or identified false positives from the \kep Object of Interest catalogue \citep[KOI,][]{2013Akeson}.
The first cross-analysis was performed using data from the National Aeronautics and Space Administration (NASA) Exoplanet Archive\footnote{See Footnote~\ref{footnote:nasa_database} for the url.}, resulting in the removal of four stars confirmed to harbour transiting planets with an orbital period matching the ones detected in our signals (see Appendix~\ref{App:transit} for an example). We also removed 59 stars identified as false positives of transiting planets.
A second comparison was carried out with the catalogue from \citet{2018Berger}, allowing for the removal of 15 stars associated with binary systems. The detected signal in the PSD is indeed most likely attributable to the binary star system. Lastly, we compared our results to the \textit{Gaia} DR3 non-single stars catalogue \citep{2022Pourbaix}, which allowed us to exclude another star from our sample, as it is part of a binary system. A total of 79 stars were removed from our sample following these cross-checks and upon completion of this step, our refined sample consists of 283 stars possibly hosting non-transiting exoplanets.

\subsection{Phase-folding, wavelet analysis, and visual inspection\label{sec:phasefold,wps}}

To further refine our study and pinpoint potential indicators of yet undetected transits, false positives, or binary systems, we phase-folded the light curves of the 283 candidates on the orbital period that we obtained from the threshold search conducted in Sect.~\ref{sec:psd}. To remove the stellar contribution at lower frequencies, we applied a finite impulse response filter \footnote{The filter is implemented through a combination of \texttt{scipy.signal.firls} and \texttt{scipy.signal.filtfit}. See preprocess function of the \texttt{star-privateer} module \citep{2024Breton}: \url{https://star-privateer.readthedocs.io/en/latest/usage/api/analysis_pipeline.html\#preprocessing}.} before phase-folding the signal.
The phase-folding analysis for KIC 5697777 is presented in panel (c) of Fig. \ref{fig:KIC_example}. The complex morphology of the phase curve suggests that different effects (see Fig.~\ref{fig:composantes}) combine to create a possible non-transiting modulation. 
A subsequent visual inspection was performed to exclude any remaining stars for which the phase-folding of the star signal fails to match the expected pattern of a non-transiting exoplanet passage (as in Fig.~\ref{fig:KIC_example}, panel (c)). As is shown in Appendix~\ref{App:phasefold}, if the trend of the phase-folded light curve differs from a stable and coherent signal over time, the candidate was discarded. This step allowed for the refinement of our overall sample to 245 candidates.\\
We also computed the wavelet power spectrum \citep[WPS,][]{Torrence1998, Liu2007} of the light curve, and we only kept stars that present stable modulations over time, as is shown in panel (d) of Fig. \ref{fig:KIC_example} for KIC 5697777. The aim is to determine whether the signal attributed to the non-transiting object exhibits a stable period throughout the different quarters of \kep’s observations. If the signal disappears during at least one quarter, this suggests that the signal may result from contamination by a nearby star. An example of a wavelet power spectrum that does not match a non-transiting signal is provided in Appendix \ref{App:wavelet}. This step removed 92 stars from our sample.\\
To ensure that the \kep signal of every star studied in this work was not contaminated by the light of a stellar neighbour with comparable brightness, we performed a search for potential contaminants using the \textit{Gaia} catalogue \citep{2023GaiaCollaboration}. We looked for \textit{Gaia} objects within 40” of each target that have an apparent magnitude in the \textit{Gaia} G band smaller than $G_{\rm candidate}+ 1$, where $G_{\rm candidate}$ is the apparent magnitude of our \kep candidate in the Gaia G band. This results in the identification of 70 Gaia objects near 38 of our targets (see Table~\ref{tb:Gaia_discardedKICs} in Appendix~\ref{App:KASOC}). For 11 of these 38 targets, the potential contaminants are located within a few arcseconds, which could result in the contamination of the \kep light curve. Consequently, we chose to remove these 11 stars from our sample.
For the remaining 27 targets, the potential contaminants are at least 10” away, making significant contamination issues unlikely. Similarly to what was done in \citet{Noraz2022}, we also performed a visual inspection on the 142 stars remaining at this stage. We individually inspected the targets using the Two Micron All Sky Survey (2MASS) images provided on the \kep Asteroseismic Science Operations Center (KASOC) database \footnote{The KASOC database can be accessed at \url{https://kasoc.phys.au.dk}.} \citep{2010Kjeldsen}. If the star is within the halo of a brighter star, it was excluded from our sample because the short-period signal might come not from a non-transiting object but rather from the nearby star in question. An example of a star whose signal is likely contaminated by a neighbouring brighter star is shown in Appendix \ref{App:KASOC}.\\
After this final step of selection, we obtained a list of 88 non-transiting exoplanet host candidates. At this point, it is important to underline that none of our candidates are in the list of 60 targets presented by \citet{Millholland2017}. We recall that they used an automated algorithm in the search for non-transiting short-period giant planets in \kep, looking for targets with an orbital period between 1 and 5 days. While they modelled the non-transiting signal and stellar variability together, we used our prior knowledge of stellar rotation to filter out the stellar contribution in the light curve. More importantly, we extended the search towards shorter orbital periods. 

\section{Fit of the photometric data\label{sec:fit of the photometric data}}

As a final step, we attempted to extract additional information on the global parameters of our candidates by fitting their signature in the light curve. Two stars were excluded from this analysis, as their signatures could not be reliably fitted in the light curve. The signal of KIC 7379583 seems contaminated by residual instrumental drifts, leading to peaks in the phase-folded light curve that prevent the fit from converging. For KIC 11350907, the signal amplitude is 84 ppm, while the random noise is around 292 ppm, making it difficult for the fit to converge.

When only photometric data were available, the resulting signal presented in Fig.~\ref{fig:composantes} depends on several free parameters, leading to degeneracies between different effects. To address this, we distinguish two cases.
On the one hand, when the phase-folding of the light curve exhibits a modulation that is approximately sinusoidal, we assumed that the non-transiting signal is dominated by emission and reflection. We then fitted only this component in the data, using the following equation \citep[see][]{Shporer2017,LilloBox2021}:

\begin{equation}
\label{eq:model1}
    F (\phi) = A_0 - A_{\rm atm} \cos (2\pi [\phi + \delta])  \; ,
\end{equation}
where $A_0$ is a flux offset parameter, and the phase, $\phi$, is given by
\begin{equation}
    \phi = (t - T_0) / P_{\rm orb} \;,
\end{equation}
where $T_0$ is taken to correspond to the time of inferior conjunction, and $\delta$ is a phase offset with respect to this conjunction. 

The amplitude, $A_{\rm atm}$, is given by
\begin{equation}
    A_{\rm atm} = 57 \; \alpha_{\mathrm{refl}} \sin i\left(\frac{M_{\star}}{M_{\odot}}\right)^{-2 / 3}\left(\frac{P}{\mathrm{day}}\right)^{-4 / 3}\left(\frac{R_{\rm p}}{R_{\mathrm{J}}}\right)^2 \, \mathrm{ppm} \; ,
    \label{eq:atm}
\end{equation}
where $\alpha_{\mathrm{refl}}$ depends on the albedo and heat recirculation efficiency. 
We adopted this approach for the 77~targets listed in Table~\ref{tab:Aatm fits}.

On the other hand, when the modulation in the phase-folding was more complex (see panel (c) of Fig.~\ref{fig:KIC_example}), we attempted to perform the fit considering the three components of the signal,

\begin{equation}
\begin{split}
\label{eq:model2}
    F (\phi) &= A_0 - A_{\rm atm} \cos (2\pi [\phi + \delta]) \\
    &- A_{\rm ellip} \cos (4\pi\phi)
    + A_{\rm boost} \sin (2\pi\phi)
    \; ,
\end{split}
\end{equation}
where $A_{\rm ellip}$ is the amplitude of the tidal deformation,
\begin{equation}
\begin{split}
    A_{\rm ellip} &= 13 \alpha_{\mathrm{ellip}} \sin i\left(\frac{R_{\star}}{R_{\odot}}\right)^3 \\
    &\times \left(\frac{M_{\star}} {M_{\odot}}\right)^{-2}\left(\frac{P}{\mathrm{day}}\right)^{-2}\left(\frac{M_{\rm p} \sin i}{M_{\mathrm{J}}}\right) \, \mathrm{ppm} \; ,
    \label{eq:ellip}
\end{split}
\end{equation}
with $\alpha_{\rm ellip}$ a coefficient that depends on the limb and gravity darkening of the host star \citep{Esteves2013}.
$A_{\rm boost}$ is the amplitude of the Doppler boosting contribution,
\begin{equation}
    A_{\rm boost} = 2.7 \; \alpha_{\mathrm{boost}}\left(\frac{P}{\mathrm{day}}\right)^{-1 / 3}\left(\frac{M_{\star}}{M_{\odot}}\right)^{-2 / 3}\left(\frac{M_{\rm p} \sin i}{M_{\mathrm{J}}}\right) \, \mathrm{ppm} \; ,
    \label{eq:boost}
\end{equation}
where $\alpha_{\mathrm{boost}}$ is the photon-weighted bandpass-integrated boosting factor. After \citet{Millholland2017}, we used a parameterisation in terms of the effective temperature, $T_{\rm eff}$, for $\alpha_{\rm ellip}$ and $\alpha_{\rm boost}$,
\begin{equation}
\begin{split}
\alpha_{\mathrm{ellip}} &= -\left(2.2 \times 10^{-4} \mathrm{~K}^{-1}\right) \; T_{\mathrm{eff}} + 2.6 \; , \\
\alpha_{\mathrm{boost}} &= - \left(6 \times 10^{-4} \mathrm{~K}^{-1}\right) \; T_{\mathrm{eff}} + 7.2 \; .
\end{split}
\end{equation}

We adopted this approach for nine targets presented in Table~\ref{tab:Agrav fits} in Appendix~\ref{App:results_fit}.

In practice, Eq.~(\ref{eq:model1}) and (\ref{eq:model2}) were fitted to the data using Bayesian inference. The log-likelihood function is
\begin{equation}
\ln \mathcal{L}_{\mathrm{\phi}}(S _{\mathrm{\phi}}, \mathrm{\theta}) = - \frac{1}{2} \times \sum_{i=1}^{n} \frac{(S _{\mathrm{\phi}}(t_i) - \mu_{\rm \phi}(\phi_i, \theta))^2}{\sigma_{\rm \phi}(t_i)^2} \; ,
\end{equation}
where $S_{\mathrm{\phi}}(t)$ is the light curve data and $\sigma_{\rm \phi}(t)$ is the normally distributed noise, considering independent errors. $\mu_{\rm \phi}(\phi, \theta)$ is the model depending on the parameter, $\theta$, as a function of the orbital phase, $\phi$, as expressed in Eq.~(\ref{eq:model1}) when following the first approach, and Eq.~(\ref{eq:model2}) when following the second approach.  We used the \texttt{UltraNest} sampler described in \cite{2021Buchner}, which is based on a nested sampling algorithm. To avoid degeneracies between $T_0$ and $\delta$, we set $\delta$ to 0 and fitted for $T_0$, $A_{\rm atm}$, and $A_0$ for the first approach, and $T_0$, $A_{\rm atm}$, $A_0$, $A_{\rm boost}$, and $A_{\rm ellip}$ for the second approach. We fixed the period, $P_{\rm orb}$, at the value inferred from the PSD of the light curve. The priors are $\mathcal{U}(-10, 10)$ for $A_0$ and $\mathcal{U}(-10^5, 10^5)$ for $A_{\mathrm{atm}}$, $A_{\mathrm{ellip}}$, and $A_{\mathrm{boost}}$, where $\mathcal{U}$(X, Y) denotes a uniform probability
distribution between the bounds X and Y. We also considered uniform priors for $T_0$ and inferred them with the timing corresponding to the minimum amplitude of the phase-folded light curves. The posterior distributions of two datasets, chosen as examples of each approach, are shown in Fig.~\ref{fig:corner_kic_4930913} and \ref{fig:corner_kic_4264634} of Appendix~\ref{App:results_fit}.

\begin{figure}[htbp!]
    \centering
    \includegraphics[width=0.5\textwidth]{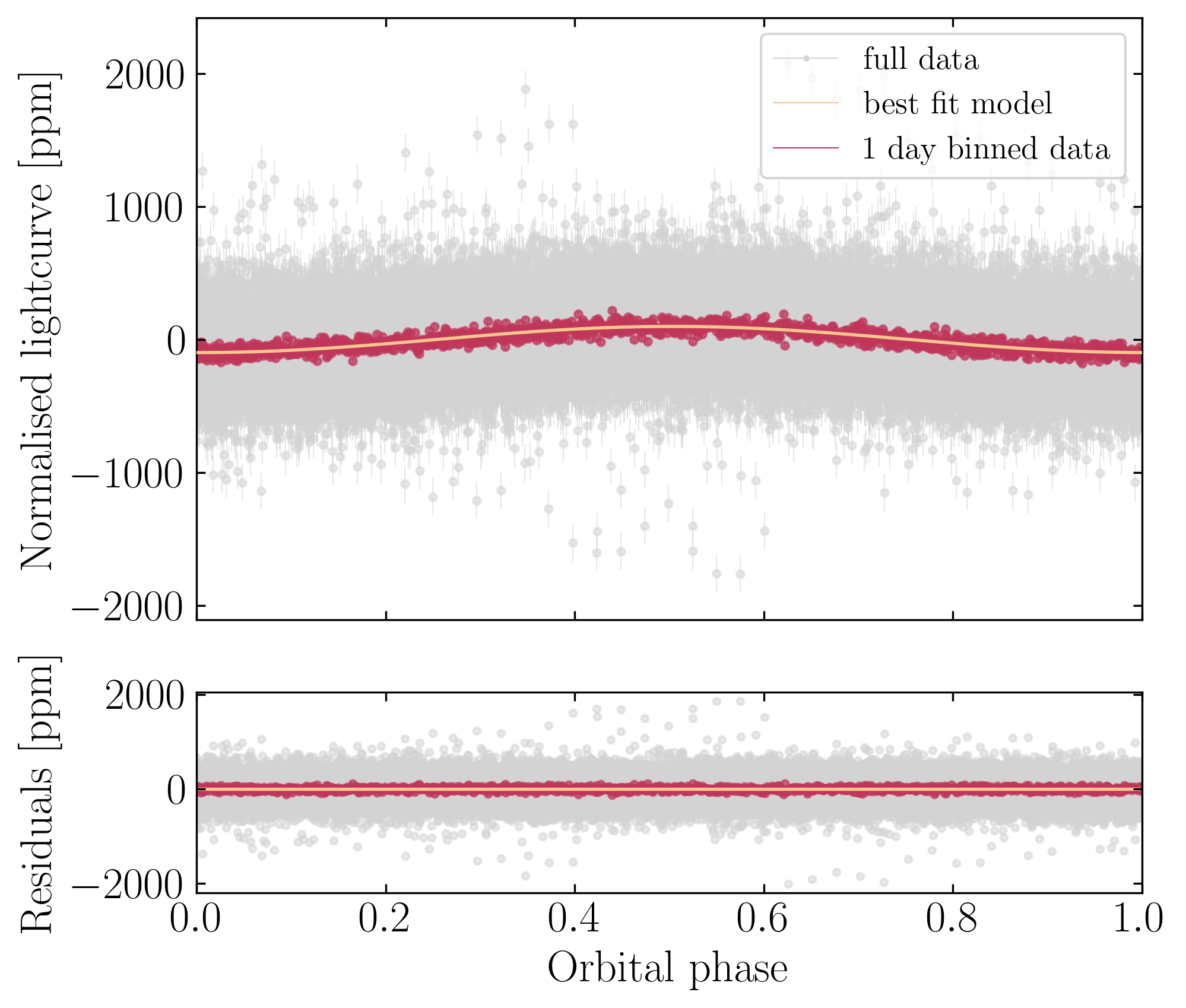}
    \caption{Phase-folded light curve of KIC~4930913, together with the  best-fit model in orange. The red points correspond to data binned in 1-day intervals. The residuals show the difference between the model and the data points and are free of large systematic trends.}
    \label{fig:bestfit_refl}
\end{figure}

At this stage, it is necessary to emphasise that the model used to perform the fit relies on a number of simplifying assumptions: the eccentricity is assumed null, the atmospheric contribution from the planet is modelled in the simplest possible way, and the stellar tidal response is considered to arise solely from equilibrium tides, with any potential contribution from dynamical tides being neglected. We therefore advise the reader to keep all these caveats in mind when they consider the results of the fit we present below. Some of the limitations of this approach are discussed in \citet{Shporer2017}.

An example of the fit result when the phase-folding of the light curve exhibits a modulation that is approximately sinusoidal is illustrated by Fig.~\ref{fig:bestfit_refl}. It shows the normalised light curve as a function of orbital phase for KIC 4930913, showing the observational data in grey, along with the best-fit model in orange. The model closely follows the general trend of the data despite significant scatter. The bottom panel displays the residuals, defined as the difference between the data and the model. The residuals are centred around zero and show no evidence of large-scale systematic trends, indicating a satisfactory fit.

\begin{figure}[htbp!]
    \begin{flushleft}
        \includegraphics[width=0.5\textwidth]{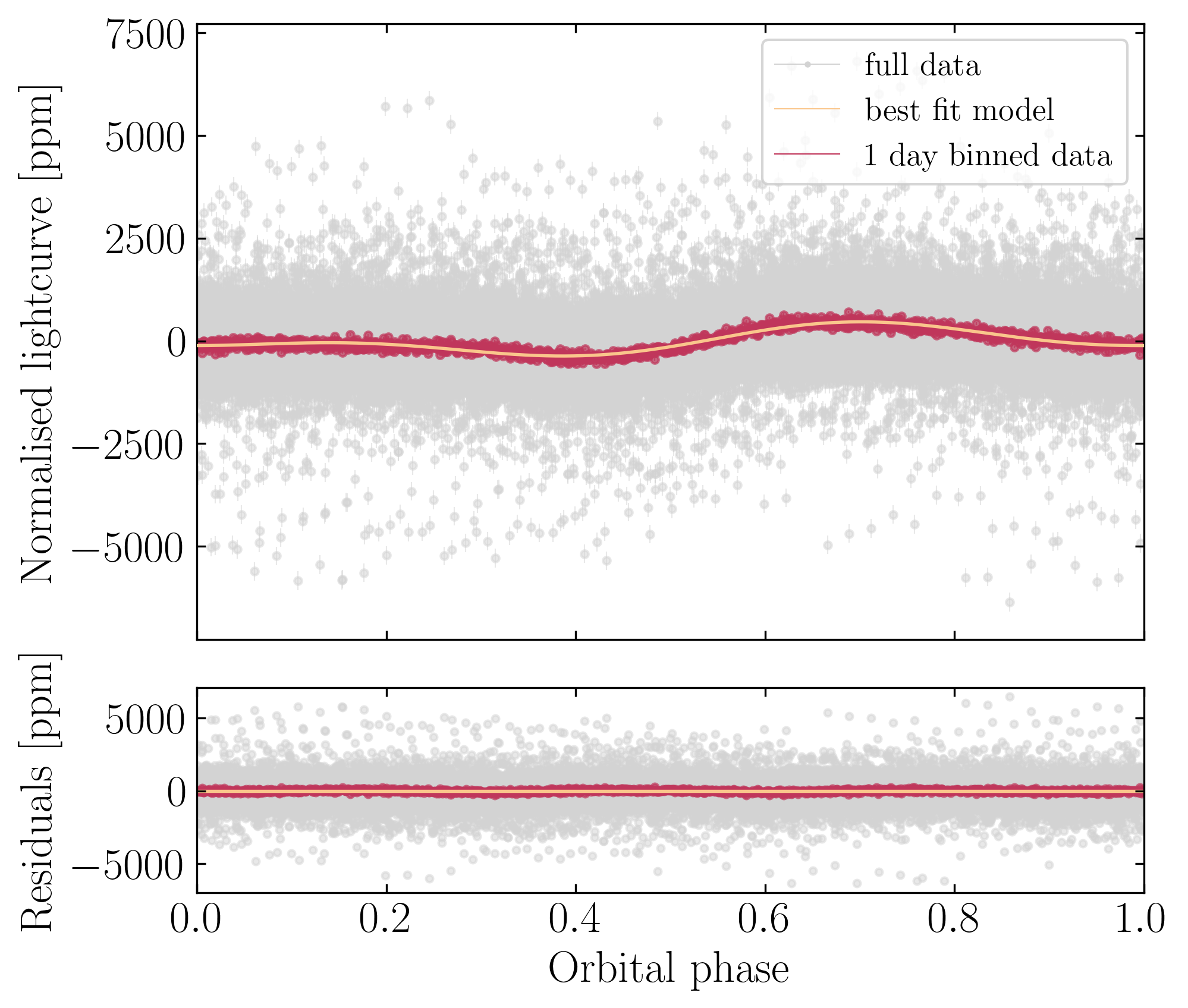}
        \caption{Phase-folded light curve of KIC~4264634, together with the  best-fit model in orange. The red points correspond to data binned in 1-day intervals. The scatter of the residuals is higher due to a larger dispersion of the data points but remains free of large systematic trends as well.}
        \label{fig:bestfit_boost_ellip}
    \end{flushleft}
\end{figure}

An example of the fit result, based on the three components of the signal described by Eq.(\ref{eq:model2}), is illustrated in Fig.\ref{fig:bestfit_boost_ellip} for KIC 4264634. The phase-folding of the light curve of this star shows a more complex pattern (in grey). As is seen in Fig.~\ref{fig:bestfit_refl}, the best-fit model (in orange) follows the general trend of the data and the residuals are again centred around zero.

The full results of the fits are displayed in Tables~\ref{tab:Aatm fits} and \ref{tab:Agrav fits}. For all 86 datasets, the posterior distribution of each parameter as well as the residuals of the fit follow a normal distribution. This indicates that the sampling successfully converged.

Using Eq.~(\ref{eq:atm}) and the fitted $A_{\rm atm}$ parameter, we derived the $\alpha_{\mathrm{refl}} \sin i
\left(R_{\rm p}/R_{\mathrm{J}}\right)^2$ parameter, obtaining a minimum value of $0.08$ and a maximum value of $21.01$ (see Table~\ref{tab:Aatm fits}).
Provided that both $\alpha_{\rm refl}$ and $i$ are known, one can then infer the radius of the object considered. Through Eq.~(\ref{eq:ellip}) and ~(\ref{eq:boost}), and the fitted parameters, $A_{\rm ellip}$ and $A_{\rm boost}$, we retrieved the $\left(M_{\rm p} \sin^{2}i/M_{\mathrm{J}}\right)$ and $\left(M_{\rm p} \sin i/M_{\mathrm{J}}\right)$ parameters, respectively. For the 77 targets analysed in the first approach, the orbital inclination is unknown and cannot be derived from the fits. However, an upper limit can be estimated by considering the geometric condition required for a transit to occur. 
For a circular orbit, a transit event occurs for i $\geq$ $i_{\rm max}$ with $i \in\left[0 ; \frac{\pi}{2}\right]$,
\begin{equation}
    i_{\rm max}=\cos ^{-1}\left(\frac{R_\star}{a}\right) \; 
    \label{eq:imax}
,\end{equation}
where $a$ is the semi-major axis of the orbit. Assuming that all of our non-transiting companions follow a circular orbit \citep{2008Mazeh}, they would then have an inclination inferior to $i_{\rm max}$. To ensure that the inclination value is in compliance with the fit, we inferred an inclination value using the ratio of $\left(M_{\rm p} \sin^{2}i/M_{\mathrm{J}}\right)$ and $\left(M_{\rm p} \sin i/M_{\mathrm{J}}\right)$, for the nine targets where $A_{\rm ellip}$ and $A_{\rm boost}$ are accessible. For three hosts (KIC~11668095, KIC~5440365, and KIC~8844761), the uncertainties of the parameters $\left(M_{\rm p} \sin^{2}i/M_{\mathrm{J}}\right)$ and $\left(M_{\rm p} \sin i/M_{\mathrm{J}}\right)$ are too large to infer $\sin i $ values. Regarding KIC~4373708, KIC~11702835, and KIC~5622796, the $\sin i$ values are inconsistent, suggesting that these stars are probably out of the range of validity of the model and would require more complex models \citep[e.g.][]{2019Penoyre}. Nevertheless, we were able to infer the inclination angle for three systems. For KIC 11550689, we obtained an inclination of 6\textdegree, for KIC 5697777, 0.4\textdegree, and for KIC 4264634, 38\textdegree. All of these values are lower than the maximum inclination angles for these systems (see Table~\ref{tab:Agrav fits}).
The corresponding masses that we can therefore derive in these three cases are 32, 1412, and 21~$\rm M_{J}$, respectively.
In the cases of KIC~11550689 and KIC~4264634, if the numerous assumptions made in this modelling approach are correct, this would suggest that the non-transiting objects might be brown dwarfs. In the case of KIC~5697777, the inferred mass is approximately 1.3~$\rm M_\odot$, which is obviously inconsistent with the hypothesis of a non-transiting substellar companion. Again, this might be because the model we apply is not valid and the inclination we obtain from this inference is not correct.  

\section{Discussion and conclusion \label{section:conclusion}}
Among our 88 candidates, we have 13 F-type, 23 G-type, 37 K-type, and 15 M-type stars exhibiting a signature consistent with the presence of a non-transiting planetary companion in a close orbit. The properties of each candidate are presented in Table \ref{tab:all_sample}. Among these candidates, we detected three systems for which the second harmonic is the largest, indicating that tidal ellipsoidal distortion outweighs the two other phenomena (see Appendix~\ref{app : 2ndharmonic systems}).

About 62\% of our sample (55 stars) exhibit orbital periods of less than a day, confirming the ability of this method to detect very short-period close-in substellar companions. In comparison, we recall again that approximately only 3\% of planets detected with the transit method exhibit orbital periods shorter than 1 day. 

Within our set of 88 candidates, two stars have been previously confirmed to harbour at least one planet, with an orbital period different from the one detected in this study (they would have been removed from our sample otherwise; see Sect.~\ref{sec:comparison with the litterature}). Details of these two candidates are provided in the Appendix \ref{appendix:appendixB}.
The method presented in this work could help increase the number of multiplanetary systems detected, which in turn can enable the scientific community to gain greater insights into the formation and evolution of exoplanetary systems.

\citet[][hereafter M13]{McQuillan2013b} identified a depletion zone for planets close to fast rotators. \citet[][hereafter G23]{Garcia2023} demonstrate that a physically motivated model that includes tidal and magnetic effects is able to reproduce the depleted region. Indeed, the amplitude of tidal effects increases dramatically as the semi-major axis gets smaller, meaning that planets in very short orbits decay very fast towards their host star when in an unstable configuration \citep{Hut1980}. In particular, the main component of the tidal potential scales as $1/a^{3}$ \citep[e.g.][]{Mathis2018}, and under some conditions the orbital decay timescale is proportional to $a^{\frac{13}{2}}$ \citep{2018Collier-Cameron}.

\begin{figure}[htbp!]
    \includegraphics[width=0.52\textwidth]{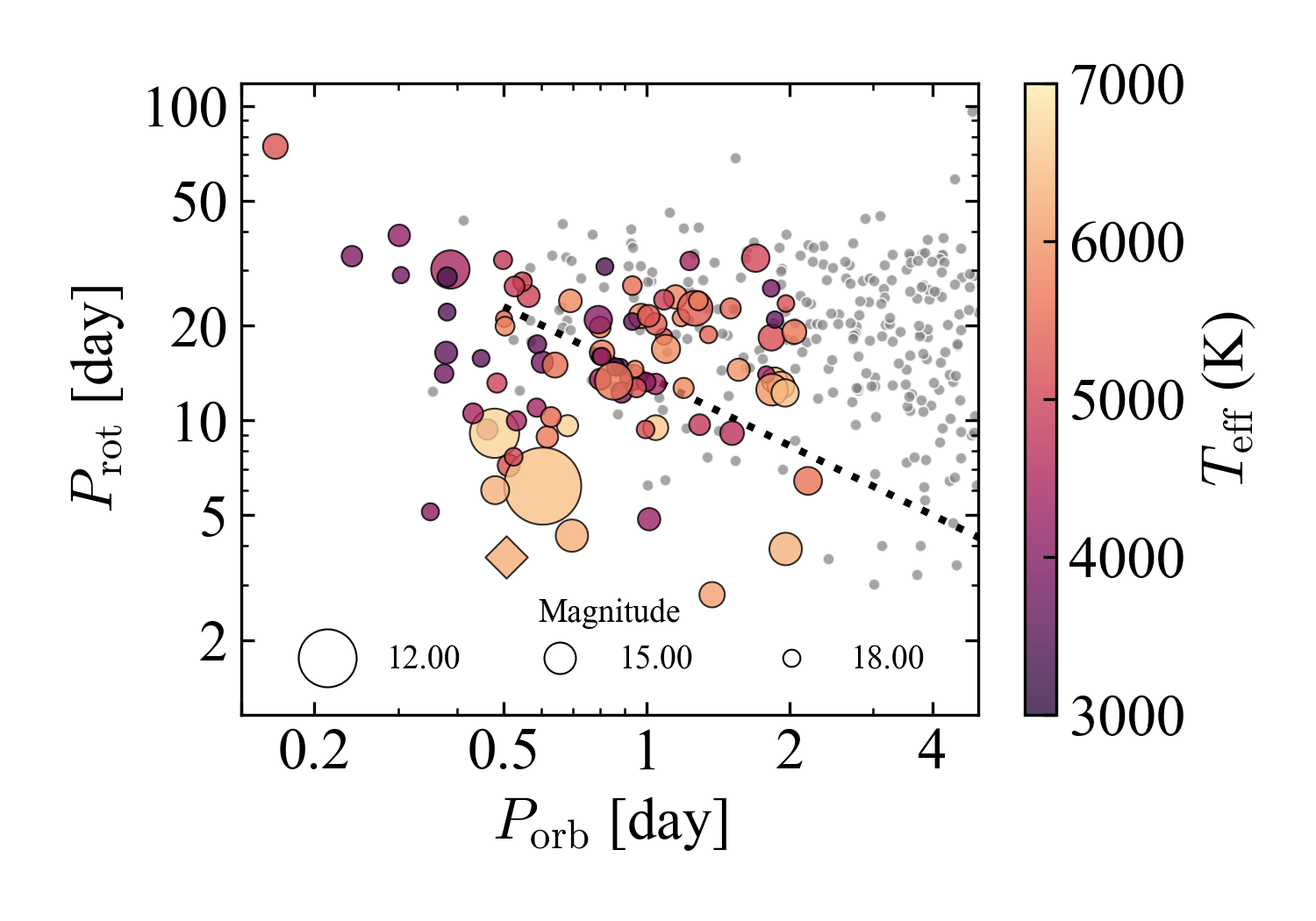}
    \caption{\prot as a function of \porb for the 88 non-transiting-exoplanet candidate systems. The size of the dots is a function of the \kep apparent magnitude of the host star, sourced from \citet{2011Brown}, whereas the temperature \teff, sourced from \citet{Berger2020}, is colour-coded. The grey sample consists of 576 confirmed single-planet-host main-sequence solar-like stars with a rotation period obtained by \citet{Garcia2023}. KIC 5697777 is represented by a diamond. The dashed black line corresponds to the fit of the lower envelope of points obtained by \citet{McQuillan2013b}.}
    \label{fig:Prot_Porb_spectraltype}
\end{figure}

Fig.~\ref{fig:Prot_Porb_spectraltype} displays the stellar rotation period \prot as a function of the orbital period \porb for our sample, linked to the findings of G23 in grey. Our 88 candidates are shown as a function of the effective temperature, \teff, sourced from \citet{Berger2020}. The dashed black line corresponds to the fit of the lower envelope of points obtained by M13.
In this context, we demonstrate that our candidates, if confirmed, would be located well below this limit, thereby populating a previously deserted area in a region where we expect exoplanets to experience orbital decay on relatively short timescales. We remind the reader that the method used in this work strongly favours the detection of planets on very short orbits. In this study, we have focused on very short orbital periods below 2.3 days. If we had been able to detect non-transiting substellar companions with frequencies exceeding our limit of 2.3 days, the region above the dashed black line would contain a higher density of candidates.

Since the orbital evolution of close-in systems takes place over short timescales, studying non-transiting exoplanets would provide us with additional systems to refine our understanding of star-planet interactions. Given the scarcity of available data on planets in short orbits around fast rotators, such a discovery would be a major asset for improving the calibration of models computing the evolution of the architecture of star-planet systems \citep[e.g.][]{Benbakoura2019,Ahuir2021}. 
In addition, by fitting phase curve models to the light curves, we have been able to extract additional constraints on the mass and radius for 86 of the detected objects. Subsequent spectroscopic ground-based follow-up is necessary to further characterise the sample identified in this work, in order to confirm the planetary nature of the candidates identified in our sample. Radial velocity observation campaigns could, for example, provide estimates of the masses of the non-transiting companions detected in this work. Such detections would be very useful to refine the set of assumptions that are used to model the signal of non-transiting planets in photometric light curves.
In the future, a similar search for non-transiting objects could be done with the K2 \citep{Howell2014}, Transiting Exoplanet Survey Satellite \citep[TESS,][]{Ricker2015}, and PLAnetary Transits and Oscillations of Stars \citep[PLATO,][]{Rauer2024} missions. Notably, the fact that PLATO will focus on bright target observations \citep{Montalto2021} will greatly expand the number of systems accessible to ground-based follow-up, allowing for the detailed characterisation of a much larger sample of exoplanetary systems.

\section*{Data avaibility}
Tables \ref{tab:Aatm fits} and \ref{tab:all_sample} are fully available in electronic form at the CDS via anonymous ftp to cdsarc.u-strasbg.fr (130.79.128.5) or via \url{http://cdsweb.u-strasbg.fr/cgi-bin/qcat?J/A+A/}.

\begin{acknowledgements}
C.G., S.N.B, and A.F.L would like to thank Dr.I. Pagano for enabling C.G.'s research internship at INAF-OACT during Spring 2023. The authors acknowledge the critical reading and the constructive comments from an anonymous referee that significantly allowed improving the original version of this paper.
A.S. and C.G. acknowledge funding from the European Research Council project ExoMagnets (grant agreement no. 101125367). A.S. acknowledges funding from the Origins project DynamEarths. A.S., R.G. and S.M. acknowledge funding from the Programme National de Planétologie (INSU/PNP).
C.G, S.N.B, and A.F.L acknowledge support from PLATO ASI-INAF agreement no. 2022-28-HH.0 "PLATO Fase D".
R.A.G acknowledges support from PLATO and SoHO/GOLF CNES grants.
A.R.G.S acknowledges the support from the FCT through national funds and FEDER through COMPETE2020 (UIDB/04434/2020, UIDP/04434/2020 \& 2022.03993.PTDC) and the support from the FCT through the work contract No. 2020.02480.CEECIND/CP1631/CT0001.
S.M.\ acknowledges support by the Spanish Ministry of Science and Innovation with  the grant no. PID2019-107061GB-C66 and through AEI under the Severo Ochoa Centres of Excellence Programme 2020--2023 (CEX2019-000920-S).
This paper includes data collected by the \kep mission, and obtained from the MAST data archive at the Space Telescope Science Institute (STScI). Funding for the \kep mission is provided by the NASA Science Mission Directorate. STScI is operated by the Association of Universities for Research in Astronomy, Inc., under NASA contract NAS 5–26555.
This research has made use of the NASA Exoplanet Archive, which is operated by the California Institute of Technology, under contract with the National Aeronautics and Space Administration under the Exoplanet Exploration Program.
\textit{Software:} 
\texttt{matplotlib} \citep{Hunter:2007}, 
\texttt{numpy} \citep{harris2020array}, 
\texttt{scipy} \citep{2020SciPy-NMeth}, 
\texttt{pandas} \citep{Pandas2020},
\texttt{pymc} \citep{pymc},
\texttt{star-privateer} \citep{2024Breton}.
\texttt{ultranest} \citep{2021Buchner}, 
\texttt{corner} \citep{corner}, 
\texttt{astropy} \citep{astropy:2022}, 
\texttt{apollinaire} \citep{Breton2022apollinaire}
\end{acknowledgements}

\bibliographystyle{aa} 
\bibliography{biblio.bib} 

\begin{thebibliography}{60}
\expandafter\ifx\csname natexlab\endcsname\relax\def\natexlab#1{#1}\fi

\bibitem[{{Ahuir} {et~al.}(2021){Ahuir}, {Strugarek}, {Brun}, \& {Mathis}}]{Ahuir2021}
{Ahuir}, J., {Strugarek}, A., {Brun}, A.~S., \& {Mathis}, S. 2021, \aap, 650, A126

\bibitem[{{Akeson} {et~al.}(2013){Akeson}, {Chen}, {Ciardi}, {Crane}, {Good}, {Harbut}, {Jackson}, {Kane}, {Laity}, {Leifer}, {Lynn}, {McElroy}, {Papin}, {Plavchan}, {Ram{\'\i}rez}, {Rey}, {von Braun}, {Wittman}, {Abajian}, {Ali}, {Beichman}, {Beekley}, {Berriman}, {Berukoff}, {Bryden}, {Chan}, {Groom}, {Lau}, {Payne}, {Regelson}, {Saucedo}, {Schmitz}, {Stauffer}, {Wyatt}, \& {Zhang}}]{2013Akeson}
{Akeson}, R.~L., {Chen}, X., {Ciardi}, D., {et~al.} 2013, \pasp, 125, 989

\bibitem[{{Appourchaux} {et~al.}(2000){Appourchaux}, {Fr{\"o}hlich}, {Andersen}, {Berthomieu}, {Chaplin}, {Elsworth}, {Finsterle}, {Gough}, {Hoeksema}, {Isaak}, {Kosovichev}, {Provost}, {Scherrer}, {Sekii}, \& {Toutain}}]{Appourchaux2000}
{Appourchaux}, T., {Fr{\"o}hlich}, C., {Andersen}, B., {et~al.} 2000, \apj, 538, 401

\bibitem[{{Astropy Collaboration} {et~al.}(2022){Astropy Collaboration}, {Price-Whelan}, {Lim}, {Earl}, {Starkman}, {Bradley}, {Shupe}, {Patil}, {Corrales}, {Brasseur}, {N{"o}the}, {Donath}, {Tollerud}, {Morris}, {Ginsburg}, {Vaher}, {Weaver}, {Tocknell}, {Jamieson}, {van Kerkwijk}, {Robitaille}, {Merry}, {Bachetti}, {G{"u}nther}, {Aldcroft}, {Alvarado-Montes}, {Archibald}, {B{'o}di}, {Bapat}, {Barentsen}, {Baz{'a}n}, {Biswas}, {Boquien}, {Burke}, {Cara}, {Cara}, {Conroy}, {Conseil}, {Craig}, {Cross}, {Cruz}, {D'Eugenio}, {Dencheva}, {Devillepoix}, {Dietrich}, {Eigenbrot}, {Erben}, {Ferreira}, {Foreman-Mackey}, {Fox}, {Freij}, {Garg}, {Geda}, {Glattly}, {Gondhalekar}, {Gordon}, {Grant}, {Greenfield}, {Groener}, {Guest}, {Gurovich}, {Handberg}, {Hart}, {Hatfield-Dodds}, {Homeier}, {Hosseinzadeh}, {Jenness}, {Jones}, {Joseph}, {Kalmbach}, {Karamehmetoglu}, {Ka{l}uszy{'n}ski}, {Kelley}, {Kern}, {Kerzendorf}, {Koch}, {Kulumani}, {Lee}, {Ly}, {Ma}, {MacBride}, {Maljaars}, {Muna}, {Murphy}, {Norman}, {O'Steen},
  {Oman}, {Pacifici}, {Pascual}, {Pascual-Granado}, {Patil}, {Perren}, {Pickering}, {Rastogi}, {Roulston}, {Ryan}, {Rykoff}, {Sabater}, {Sakurikar}, {Salgado}, {Sanghi}, {Saunders}, {Savchenko}, {Schwardt}, {Seifert-Eckert}, {Shih}, {Jain}, {Shukla}, {Sick}, {Simpson}, {Singanamalla}, {Singer}, {Singhal}, {Sinha}, {Sip{H{o}}cz}, {Spitler}, {Stansby}, {Streicher}, {{{S}}umak}, {Swinbank}, {Taranu}, {Tewary}, {Tremblay}, {Val-Borro}, {Van Kooten}, {Vasovi{'c}}, {Verma}, {de Miranda Cardoso}, {Williams}, {Wilson}, {Winkel}, {Wood-Vasey}, {Xue}, {Yoachim}, {Zhang}, {Zonca}, \& {Astropy Project Contributors}}]{astropy:2022}
{Astropy Collaboration}, {Price-Whelan}, A.~M., {Lim}, P.~L., {et~al.} 2022, apj, 935, 167

\bibitem[{{Barker} {et~al.}(2024){Barker}, {Efroimsky}, {Makarov}, \& {Veras}}]{2024Barker}
{Barker}, A.~J., {Efroimsky}, M., {Makarov}, V.~V., \& {Veras}, D. 2024, \mnras, 527, 5131

\bibitem[{{Benbakoura} {et~al.}(2019){Benbakoura}, {R{\'e}ville}, {Brun}, {Le Poncin-Lafitte}, \& {Mathis}}]{Benbakoura2019}
{Benbakoura}, M., {R{\'e}ville}, V., {Brun}, A.~S., {Le Poncin-Lafitte}, C., \& {Mathis}, S. 2019, \aap, 621, A124

\bibitem[{{Berger} {et~al.}(2018){Berger}, {Huber}, {Gaidos}, \& {van Saders}}]{2018Berger}
{Berger}, T.~A., {Huber}, D., {Gaidos}, E., \& {van Saders}, J.~L. 2018, \apj, 866, 99

\bibitem[{{Berger} {et~al.}(2020){Berger}, {Huber}, {van Saders}, {Gaidos}, {Tayar}, \& {Kraus}}]{Berger2020}
{Berger}, T.~A., {Huber}, D., {van Saders}, J.~L., {et~al.} 2020, \aj, 159, 280

\bibitem[{{Borucki} {et~al.}(2010){Borucki}, {Koch}, {Basri}, {Batalha}, {Brown}, {Caldwell}, {Caldwell}, {Christensen-Dalsgaard}, {Cochran}, {DeVore}, {Dunham}, {Dupree}, {Gautier}, {Geary}, {Gilliland}, {Gould}, {Howell}, {Jenkins}, {Kondo}, {Latham}, {Marcy}, {Meibom}, {Kjeldsen}, {Lissauer}, {Monet}, {Morrison}, {Sasselov}, {Tarter}, {Boss}, {Brownlee}, {Owen}, {Buzasi}, {Charbonneau}, {Doyle}, {Fortney}, {Ford}, {Holman}, {Seager}, {Steffen}, {Welsh}, {Rowe}, {Anderson}, {Buchhave}, {Ciardi}, {Walkowicz}, {Sherry}, {Horch}, {Isaacson}, {Everett}, {Fischer}, {Torres}, {Johnson}, {Endl}, {MacQueen}, {Bryson}, {Dotson}, {Haas}, {Kolodziejczak}, {Van Cleve}, {Chandrasekaran}, {Twicken}, {Quintana}, {Clarke}, {Allen}, {Li}, {Wu}, {Tenenbaum}, {Verner}, {Bruhweiler}, {Barnes}, \& {Prsa}}]{Borucki2010}
{Borucki}, W.~J., {Koch}, D., {Basri}, G., {et~al.} 2010, Science, 327, 977

\bibitem[{{Breton} {et~al.}(2022){Breton}, {Garc{\'\i}a}, {Ballot}, {Delsanti}, \& {Salabert}}]{Breton2022apollinaire}
{Breton}, S.~N., {Garc{\'\i}a}, R.~A., {Ballot}, J., {Delsanti}, V., \& {Salabert}, D. 2022, \aap, 663, A118

\bibitem[{{Breton} {et~al.}(2024){Breton}, {Lanza}, {Messina}, {Pagano}, {Bugnet}, {Corsaro}, {Garc{'i}a}, {Mathur}, {Santos}, {Aigrain}, {Amard}, {Brun}, {Degott}, {Noraz}, {Palakkatharappil}, {Panetier}, {Strugarek}, {Belkacem}, {Goupil}, {Ouazzani}, {Philidet}, {Reni{'e}}, \& {Roth}}]{2024Breton}
{Breton}, S.~N., {Lanza}, A.~F., {Messina}, S., {et~al.} 2024, \aap, 689, A229

\bibitem[{{Brown} {et~al.}(2011){Brown}, {Latham}, {Everett}, \& {Esquerdo}}]{2011Brown}
{Brown}, T.~M., {Latham}, D.~W., {Everett}, M.~E., \& {Esquerdo}, G.~A. 2011, \aj, 142, 112

\bibitem[{{Buchner}(2021)}]{2021Buchner}
{Buchner}, J. 2021, The Journal of Open Source Software, 6, 3001

\bibitem[{{Collier Cameron} \& {Jardine}(2018)}]{2018Collier-Cameron}
{Collier Cameron}, A. \& {Jardine}, M. 2018, \mnras, 476, 2542

\bibitem[{{Cullen} \& {Bayliss}(2024)}]{2024Cullen}
{Cullen}, C.~J. \& {Bayliss}, D. 2024, \mnras, 531, 1133

\bibitem[{Esteves {et~al.}(2013)Esteves, De~Mooij, \& Jayawardhana}]{Esteves2013}
Esteves, L.~J., De~Mooij, E. J.~W., \& Jayawardhana, R. 2013, The Astrophysical Journal, 772, 51

\bibitem[{Foreman-Mackey(2016)}]{corner}
Foreman-Mackey, D. 2016, The Journal of Open Source Software, 1, 24

\bibitem[{Fortney {et~al.}(2021)Fortney, Dawson, \& Komacek}]{Fortney2021}
Fortney, J.~J., Dawson, R.~I., \& Komacek, T.~D. 2021, Journal of Geophysical Research: Planets, 126, e2020JE006629, \_eprint: https://onlinelibrary.wiley.com/doi/pdf/10.1029/2020JE006629

\bibitem[{{Gaia Collaboration} {et~al.}(2023){Gaia Collaboration}, {Vallenari}, {Brown}, {Prusti}, {de Bruijne}, {Arenou}, {Babusiaux}, {Biermann}, {Creevey}, {Ducourant}, {Evans}, {Eyer}, {Guerra}, {Hutton}, {Jordi}, {Klioner}, {Lammers}, {Lindegren}, {Luri}, {Mignard}, {Panem}, {Pourbaix}, {Randich}, {Sartoretti}, {Soubiran}, {Tanga}, {Walton}, {Bailer-Jones}, {Bastian}, {Drimmel}, {Jansen}, {Katz}, {Lattanzi}, {van Leeuwen}, {Bakker}, {Cacciari}, {Casta{\~n}eda}, {De Angeli}, {Fabricius}, {Fouesneau}, {Fr{\'e}mat}, {Galluccio}, {Guerrier}, {Heiter}, {Masana}, {Messineo}, {Mowlavi}, {Nicolas}, {Nienartowicz}, {Pailler}, {Panuzzo}, {Riclet}, {Roux}, {Seabroke}, {Sordo}, {Th{\'e}venin}, {Gracia-Abril}, {Portell}, {Teyssier}, {Altmann}, {Andrae}, {Audard}, {Bellas-Velidis}, {Benson}, {Berthier}, {Blomme}, {Burgess}, {Busonero}, {Busso}, {C{\'a}novas}, {Carry}, {Cellino}, {Cheek}, {Clementini}, {Damerdji}, {Davidson}, {de Teodoro}, {Nu{\~n}ez Campos}, {Delchambre}, {Dell'Oro}, {Esquej},
  {Fern{\'a}ndez-Hern{\'a}ndez}, {Fraile}, {Garabato}, {Garc{\'\i}a-Lario}, {Gosset}, {Haigron}, {Halbwachs}, {Hambly}, {Harrison}, {Hern{\'a}ndez}, {Hestroffer}, {Hodgkin}, {Holl}, {Jan{\ss}en}, {Jevardat de Fombelle}, {Jordan}, {Krone-Martins}, {Lanzafame}, {L{\"o}ffler}, {Marchal}, {Marrese}, {Moitinho}, {Muinonen}, {Osborne}, {Pancino}, {Pauwels}, {Recio-Blanco}, {Reyl{\'e}}, {Riello}, {Rimoldini}, {Roegiers}, {Rybizki}, {Sarro}, {Siopis}, {Smith}, {Sozzetti}, {Utrilla}, {van Leeuwen}, {Abbas}, {{\'A}brah{\'a}m}, {Abreu Aramburu}, {Aerts}, {Aguado}, {Ajaj}, {Aldea-Montero}, {Altavilla}, {{\'A}lvarez}, {Alves}, {Anders}, {Anderson}, {Anglada Varela}, {Antoja}, {Baines}, {Baker}, {Balaguer-N{\'u}{\~n}ez}, {Balbinot}, {Balog}, {Barache}, {Barbato}, {Barros}, {Barstow}, {Bartolom{\'e}}, {Bassilana}, {Bauchet}, {Becciani}, {Bellazzini}, {Berihuete}, {Bernet}, {Bertone}, {Bianchi}, {Binnenfeld}, {Blanco-Cuaresma}, {Blazere}, {Boch}, {Bombrun}, {Bossini}, {Bouquillon}, {Bragaglia}, {Bramante}, {Breedt},
  {Bressan}, {Brouillet}, {Brugaletta}, {Bucciarelli}, {Burlacu}, {Butkevich}, {Buzzi}, {Caffau}, {Cancelliere}, {Cantat-Gaudin}, {Carballo}, {Carlucci}, {Carnerero}, {Carrasco}, {Casamiquela}, {Castellani}, {Castro-Ginard}, {Chaoul}, {Charlot}, {Chemin}, {Chiaramida}, {Chiavassa}, {Chornay}, {Comoretto}, {Contursi}, {Cooper}, {Cornez}, {Cowell}, {Crifo}, {Cropper}, {Crosta}, {Crowley}, {Dafonte}, {Dapergolas}, {David}, {David}, {de Laverny}, {De Luise}, \& {De March}}]{2023GaiaCollaboration}
{Gaia Collaboration}, {Vallenari}, A., {Brown}, A.~G.~A., {et~al.} 2023, \aap, 674, A1

\bibitem[{{Garc{\'\i}a} \& {Ballot}(2019)}]{Garcia2019}
{Garc{\'\i}a}, R.~A. \& {Ballot}, J. 2019, Living Reviews in Solar Physics, 16, 4

\bibitem[{{Garc{\'\i}a} {et~al.}(2023){Garc{\'\i}a}, {Gourv{\`e}s}, {Santos}, {Strugarek}, {Godoy-Rivera}, {Mathur}, {Delsanti}, {Breton}, {Beck}, {Brun}, \& {Mathis}}]{Garcia2023}
{Garc{\'\i}a}, R.~A., {Gourv{\`e}s}, C., {Santos}, {\^A}.~R.~G., {et~al.} 2023, \aap, 679, L12

\bibitem[{{Garc{\'\i}a} {et~al.}(2011){Garc{\'\i}a}, {Hekker}, {Stello}, {Guti{\'e}rrez-Soto}, {Handberg}, {Huber}, {Karoff}, {Uytterhoeven}, {Appourchaux}, {Chaplin}, {Elsworth}, {Mathur}, {Ballot}, {Christensen-Dalsgaard}, {Gilliland}, {Houdek}, {Jenkins}, {Kjeldsen}, {McCauliff}, {Metcalfe}, {Middour}, {Molenda-Zakowicz}, {Monteiro}, {Smith}, \& {Thompson}}]{Garcia2011}
{Garc{\'\i}a}, R.~A., {Hekker}, S., {Stello}, D., {et~al.} 2011, \mnras, 414, L6

\bibitem[{{Garc{\'\i}a} {et~al.}(2014){Garc{\'\i}a}, {Mathur}, {Pires}, {R{\'e}gulo}, {Bellamy}, {Pall{\'e}}, {Ballot}, {Barcel{\'o} Forteza}, {Beck}, \& {Bedding}}]{Garcia2014}
{Garc{\'\i}a}, R.~A., {Mathur}, S., {Pires}, S., {et~al.} 2014, \aap, 568, A10

\bibitem[{Harris {et~al.}(2020)Harris, Millman, van~der Walt, Gommers, Virtanen, Cournapeau, Wieser, Taylor, Berg, Smith, Kern, Picus, Hoyer, van Kerkwijk, Brett, Haldane, del R{\'{i}}o, Wiebe, Peterson, G{\'{e}}rard-Marchant, Sheppard, Reddy, Weckesser, Abbasi, Gohlke, \& Oliphant}]{harris2020array}
Harris, C.~R., Millman, K.~J., van~der Walt, S.~J., {et~al.} 2020, Nature, 585, 357

\bibitem[{{Howell} {et~al.}(2014){Howell}, {Sobeck}, {Haas}, {Still}, {Barclay}, {Mullally}, {Troeltzsch}, {Aigrain}, {Bryson}, {Caldwell}, {Chaplin}, {Cochran}, {Huber}, {Marcy}, {Miglio}, {Najita}, {Smith}, {Twicken}, \& {Fortney}}]{Howell2014}
{Howell}, S.~B., {Sobeck}, C., {Haas}, M., {et~al.} 2014, \pasp, 126, 398

\bibitem[{Hunter(2007)}]{Hunter:2007}
Hunter, J.~D. 2007, Computing in Science \& Engineering, 9, 90

\bibitem[{{Hut}(1980)}]{Hut1980}
{Hut}, P. 1980, \aap, 92, 167

\bibitem[{{Kjeldsen} {et~al.}(2010){Kjeldsen}, {Christensen-Dalsgaard}, {Handberg}, {Brown}, {Gilliland}, {Borucki}, \& {Koch}}]{2010Kjeldsen}
{Kjeldsen}, H., {Christensen-Dalsgaard}, J., {Handberg}, R., {et~al.} 2010, Astronomische Nachrichten, 331, 966

\bibitem[{{Lillo-Box} {et~al.}(2021){Lillo-Box}, {Millholland}, \& {Laughlin}}]{LilloBox2021}
{Lillo-Box}, J., {Millholland}, S., \& {Laughlin}, G. 2021, \aap, 654, A9

\bibitem[{Liu {et~al.}(2007)Liu, San~Liang, \& Weisberg}]{Liu2007}
Liu, Y., San~Liang, X., \& Weisberg, R.~H. 2007, Journal of Atmospheric and Oceanic Technology, 24, 2093

\bibitem[{{Mathis}(2018)}]{Mathis2018}
{Mathis}, S. 2018, in Handbook of Exoplanets, ed. H.~J. {Deeg} \& J.~A. {Belmonte}, 24

\bibitem[{{Mathur} {et~al.}(2017){Mathur}, {Huber}, {Batalha}, {Ciardi}, {Bastien}, {Bieryla}, {Buchhave}, {Cochran}, {Endl}, {Esquerdo}, {Furlan}, {Howard}, {Howell}, {Isaacson}, {Latham}, {MacQueen}, \& {Silva}}]{Mathur2017}
{Mathur}, S., {Huber}, D., {Batalha}, N.~M., {et~al.} 2017, \apjs, 229, 30

\bibitem[{{Mayor} \& {Queloz}(1995)}]{Mayor1995}
{Mayor}, M. \& {Queloz}, D. 1995, \nat, 378, 355

\bibitem[{{Mazeh}(2008)}]{2008Mazeh}
{Mazeh}, T. 2008, in EAS Publications Series, Vol.~29, EAS Publications Series, ed. M.~J. {Goupil} \& J.~P. {Zahn}, 1--65

\bibitem[{{McQuillan} {et~al.}(2013){McQuillan}, {Mazeh}, \& {Aigrain}}]{McQuillan2013b}
{McQuillan}, A., {Mazeh}, T., \& {Aigrain}, S. 2013, \apjl, 775, L11

\bibitem[{{Messias} {et~al.}(2022){Messias}, {de Oliveira}, {Gomes}, {Arruda Gon{\c{c}}alves}, {Canto Martins}, {Le{\~a}o}, \& {De Medeiros}}]{Messias2022}
{Messias}, Y.~S., {de Oliveira}, L.~L.~A., {Gomes}, R.~L., {et~al.} 2022, \apjl, 930, L23

\bibitem[{{Millholland} \& {Laughlin}(2017)}]{Millholland2017}
{Millholland}, S. \& {Laughlin}, G. 2017, \aj, 154, 83

\bibitem[{{Montalto} {et~al.}(2021){Montalto}, {Piotto}, {Marrese}, {Nascimbeni}, {Prisinzano}, {Granata}, {Marinoni}, {Desidera}, {Ortolani}, {Aerts}, {Alei}, {Altavilla}, {Benatti}, {B{\"o}rner}, {Cabrera}, {Claudi}, {Deleuil}, {Fabrizio}, {Gizon}, {Goupil}, {Heras}, {Magrin}, {Malavolta}, {Mas-Hesse}, {Pagano}, {Paproth}, {Pertenais}, {Pollacco}, {Ragazzoni}, {Ramsay}, {Rauer}, \& {Udry}}]{Montalto2021}
{Montalto}, M., {Piotto}, G., {Marrese}, P.~M., {et~al.} 2021, \aap, 653, A98

\bibitem[{{Morton} {et~al.}(2016){Morton}, {Bryson}, {Coughlin}, {Rowe}, {Ravichandran}, {Petigura}, {Haas}, \& {Batalha}}]{2016Morton}
{Morton}, T.~D., {Bryson}, S.~T., {Coughlin}, J.~L., {et~al.} 2016, \apj, 822, 86

\bibitem[{{Noraz} {et~al.}(2022){Noraz}, {Breton}, {Brun}, {Garc{\'\i}a}, {Strugarek}, {Santos}, {Mathur}, \& {Amard}}]{Noraz2022}
{Noraz}, Q., {Breton}, S.~N., {Brun}, A.~S., {et~al.} 2022, \aap

\bibitem[{pandas~development team(2020)}]{Pandas2020}
pandas~development team, T. 2020, pandas-dev/pandas: Pandas

\bibitem[{{Patra} {et~al.}(2017){Patra}, {Winn}, {Holman}, {Yu}, {Deming}, \& {Dai}}]{Patra2017}
{Patra}, K.~C., {Winn}, J.~N., {Holman}, M.~J., {et~al.} 2017, \aj, 154, 4

\bibitem[{{Penoyre} \& {Sandford}(2019)}]{2019Penoyre}
{Penoyre}, Z. \& {Sandford}, E. 2019, \mnras, 488, 4181

\bibitem[{{Pires} {et~al.}(2015){Pires}, {Mathur}, {Garc{\'\i}a}, {Ballot}, {Stello}, \& {Sato}}]{Pires2015}
{Pires}, S., {Mathur}, S., {Garc{\'\i}a}, R.~A., {et~al.} 2015, \aap, 574, A18

\bibitem[{{Pourbaix} {et~al.}(2022){Pourbaix}, {Arenou}, {Gavras}, {Gosset}, {Halbwachs}, {Siopis}, {Sozzetti}, {Bauchet}, {Damerdji}, {Delchambre}, {Delisle}, {Giacobbe}, {Holl}, {Jorissen}, {Lattanzi}, {Leclerc}, {Morel}, {Sadowski}, {Sahlmann}, \& {Segransan}}]{2022Pourbaix}
{Pourbaix}, D., {Arenou}, F., {Gavras}, P., {et~al.} 2022, {Gaia DR3 documentation Chapter 7: Non-single stars}

\bibitem[{{Rauer} {et~al.}(2024){Rauer}, {Aerts}, {Cabrera}, {Deleuil}, {Erikson}, {Gizon}, {Goupil}, {Heras}, {Lorenzo-Alvarez}, {Marliani}, {Martin-Garcia}, {Mas-Hesse}, {O'Rourke}, {Osborn}, {Pagano}, {Piotto}, {Pollacco}, {Ragazzoni}, {Ramsay}, {Udry}, {Appourchaux}, {Benz}, {Brandeker}, {G{\"u}del}, {Janot-Pacheco}, {Kabath}, {Kjeldsen}, {Min}, {Santos}, {Smith}, {Suarez}, {Werner}, {Aboudan}, {Abreu}, {Acu a}, {Adams}, {Adibekyan}, {Affer}, {Agneray}, {Agnor}, {Aguirre B{\o}rsen-Koch}, {Ahmed}, {Aigrain}, {Al-Bahlawan}, {Alcacera Gil}, {Alei}, {Alencar}, {Alexander}, {Alfonso-Garz{\'o}n}, {Alibert}, {Allende Prieto}, {Almeida}, {Alonso Sobrino}, {Altavilla}, {Althaus}, {Alonso Alvarez Trujillo}, {Amarsi}, {Ammler-von Eiff}, {Am{\^o}res}, {Andrade}, {Antoniadis-Karnavas}, {Ant{\'o}nio}, {Aparicio del Moral}, {Appolloni}, {Arena}, {Armstrong}, {Aroca Aliaga}, {Asplund}, {Audenaert}, {Auricchio}, {Avelino}, {Baeke}, {Bailli{\'e}}, {Balado}, {Ballber Balaguer{\'o}}, {Balestra}, {Ball}, {Ballans}, {Ballot},
  {Barban}, {Barbary}, {Barbieri}, {Barcel{\'o} Forteza}, {Barker}, {Barklem}, {Barnes}, {Barrado Navascues}, {Barragan}, {Baruteau}, {Basu}, {Baudin}, {Baumeister}, {Bayliss}, {Bazot}, {Beck}, {Bedding}, {Belkacem}, {Bellinger}, {Benatti}, {Benomar}, {B{\'e}rard}, {Bergemann}, {Bergomi}, {Bernardo}, {Biazzo}, {Bignamini}, {Bigot}, {Billot}, {Binet}, {Biondi}, {Biondi}, {Birch}, {Bitsch}, {Bluhm Ceballos}, {B{\'o}di}, {Bogn{\'a}r}, {Boisse}, {Bolmont}, {Bonanno}, {Bonavita}, {Bonfanti}, {Bonfils}, {Bonito}, {Bonomo}, {B{\"o}rner}, {Boro Saikia}, {Borreguero Mart{\'\i}n}, {Borsa}, {Borsato}, {Bossini}, {Bouchy}, {Bou{\'e}}, {Boufleur}, {Boumier}, {Bourrier}, {Bowman}, {Bozzo}, {Bradley}, {Bray}, {Bressan}, {Breton}, {Brienza}, {Brito}, {Brogi}, {Brown}, {Brown}, {Brun}, {Bruno}, {Bruns}, {Buchhave}, {Bugnet}, {Buldgen}, {Burgess}, {Busatta}, {Busso}, {Buzasi}, {Caballero}, {Cabral}, {Cabrero Gomez}, {Calderone}, {Cameron}, {Cameron}, {Campante}, {Campos Gestal}, {Canto Martins}, {Cara}, {Carone}, {Carrasco},
  {Casagrande}, {Casewell}, {Cassisi}, {Castellani}, {Castro}, {Catala}, {Catal{\'a}n Fern{\'a}ndez}, {Catelan}, {Cegla}, {Cerruti}, {Cessa}, {Chadid}, {Chaplin}, {Charpinet}, {Chiappini}, {Chiarucci}, {Chiavassa}, {Chinellato}, {Chirulli}, {Christensen-Dalsgaard}, {Church}, {Claret}, {Clarke}, {Claudi}, {Clermont}, {Coelho}, {Coelho}, {Cogato}, {Colom{\'e}}, {Condamin}, {Conde Garc{\'\i}a}, {Conseil}, {Corbard}, {Correia}, {Corsaro}, {Cosentino}, {Costes}, {Cottinelli}, {Covone}, {Creevey}, {Crida}, {Csizmadia}, {Cunha}, {Curry}, {da Costa}, {da Silva}, {Dalal}, {Damasso}, {Damiani}, {Damiani}, {Liduina das Chagas}, {Davies}, {Davies}, {Davies}, {Davison}, {de Almeida}, {de Angeli}, {Cabral de Barros}, {de Castro Le{\~a}o}, {Brito de Freitas}, {de Freitas}, {De Martino}, {Renan de Medeiros}, {de Paula}, {de Pedraza G{\'o}mez}, {de Plaa}, {De Ridder}, {Deal}, {Decin}, {Deeg}, {Degl Innocenti}, {Deheuvels}, {del Burgo}, {Del Sordo}, {Delgado-Mena}, {Demangeon}, {Denk}, {Derekas}, {Desert}, {Desidera}, {Dexet},
  {Di Criscienzo}, {Di Giorgio}, {Di Mauro}, {Diaz Rial}, {D{\'\i}az-Garc{\'\i}a}, {Dima}, {Dinuzzi}, {Dionatos}, {Distefano}, {do Nascimento}, {Domingo}, {D'Orazi}, {Dorn}, {Doyle}, {Duarte}, {Ducellier}, {Dumaye}, {Dumusque}, {Dupret}, {Eggenberger}, {Ehrenreich}, {Eigm{\"u}ller}, {Eising}, {Emilio}, {Eriksson}, {Ermocida}, {Isidoro Escate Giribaldi}, {Eschen}, {ez}, {Estrela}, {Evans}, {Fabbian}, {Fabrizio}, {Faria}, {Farina}, {Farinato}, {Feliz}, {Feltzing}, {Fenouillet}, {Fern{\'a}ndez}, {Ferrari}, {Ferraz-Mello}, {Fialho}, {Fienga}, {Figueira}, {Fiori}, {Flaccomio}, {Focardi}, {Foley}, {Fontignie}, {Ford}, {Fornazier}, {Forveille}, {Fossati}, {de Marca Franca}, {da Silva}, {Frasca}, {Fridlund}, {Furlan}, {Gabler}, {Gaido}, {Gallagher}, {Gallego Sempere}, {Galli}, {Garc{\'\i}a}, {Garc{\'\i}a Hern{\'a}ndez}, {Garcia Munoz}, {Garc{\'\i}a-V{\'a}zquez}, {Garrido Haba}, {Gaulme}, {Gauthier}, {Gehan}, {Gent}, {Georgieva}, {Ghigo}, {Giana}, {Gill}, {Girardi}, {Giuliatti Winter}, {Giusi}, {Gomes da Silva},
  {G{\'o}mez Zazo}, {Gomez-Lopez}, {Isai Gonz{\'a}lez Hern{\'a}ndez}, {Gonzalez Murillo}, {Gonzalo Melchor}, {Gorius}, {Gouel}, {Goulty}, {Granata}, {Grenfell}, {Grie bach}, {Grolleau}, {Grouffal}, {Grziwa}, {Guarcello}, {Gueguen}, {Guenther}, {Guilhem}, {Guillerot}, {Guillot}, {Guiot}, {Guterman}, {Guti{\'e}rrez}, {Guti{\'e}rrez-Canales}, {Hagelberg}, {Haldemann}, {Hall}, {Handberg}, {Harrison}, {Harrison}, {Hasiba}, {Haswell}, {Hatalova}, {Hatzes}, {Haywood}, {H{\'e}brard}, {Heckes}, {Heiter}, {Hekker}, {Heller}, {Helling}, {Helminiak}, {Hemsley}, {Heng}, {Herbst}, {Hermans}, {Hermes}, {Hidalgo Torres}, {Hinkel}, {Hobbs}, {Hodgkin}, {Hofmann}, {Hojjatpanah}, {Houdek}, {Huber}, {Huesler}, {Hui-Bon-Hoa}, {Huygen}, {Huynh}, {Iro}, {Irwin}, {Irwin}, {Izidoro}, {Jacquinod}, {Emborg Jannsen}, {Janson}, {Jeszenszky}, {Jiang}, {Jos{\'e} Jimenez Mancebo}, {Jofre}, {Johansen}, {Johnston}, {Jones}, {Kallinger}, {K{\'a}lm{\'a}n}, {Kanitz}, {Karjalainen}, {Karjalainen}, {Karoff}, {Kawaler}, {Kawata}, {Keereman},
  {Keiderling}, {Kennedy}, {Kenworthy}, {Kerschbaum}, {Kidger}, {Kiefer}, {Kintziger}, {Kislyakova}, {Kiss}, {Klagyivik}, {Klahr}, {Klevas}, {Kochukhov}, {K{\"o}hler}, {Kolb}, {Koncz}, {Korth}, {Kostogryz}, {Kov{\'a}cs}, {Kov{\'a}cs}, {Kozhura}, {Krivova}, {Kucinskas}, {Kuhlemann}, {Kupka}, {Laauwen}, {Labiano}, {Lagarde}, {Laget}, {Laky}, {Lam}, {Lambrechts}, {Lammer}, {Lanza}, {Lanzafame}, {Lares Martiz}, {Laskar}, {Latter}, {Lavanant}, {Lawrenson}, {Lazzoni}, {Lebre}, {Lebreton}, {Lecavelier des Etangs}, {Lee}, {Leinhardt}, {Leleu}, {Lendl}, {Leto}, {Levillain}, {Libert}, {Lichtenberg}, {Ligi}, {Lignieres}, {Lillo-Box}, {Linsky}, {Scige Liu}, {Loidolt}, {Longval}, {Lopes}, {Lorenzani}, {Ludwig}, {Lund}, {Sloth Lundkvist}, {Luri}, {Maceroni}, {Madden}, {Madhusudhan}, {Maggio}, {Magliano}, {Magrin}, {Mahy}, {Maibaum}, {Malac-Allain}, {Malapert}, {Malavolta}, {Maldonado}, {Mamonova}, {Manchon}, {Manj{\'o}n}, {Mann}, {Mantovan}, {Marafatto}, {Marconi}, {Mardling}, {Marigo}, {Marinoni}, {Marques}, {Marques},
  {Marrese}, {Marshall}, {Mart{\'\i}nez Perales}, {Mary}, {Marzari}, {Masana}, {Mascher}, {Mathis}, {Mathur}, {Mart{\'\i}n Vodopivec}, {Mattiuci Figueiredo}, {Maxted}, {Mazeh}, {Mazevet}, {Mazzei}, {McCormac}, {McMillan}, {Menou}, {Merle}, {Meru}, {Mesa}, {Messina}, {M{\'e}sz{\'a}ros}, {Meunier}, {Meunier}, {Micela}, {Michaelis}, {Michel}, {Michielsen}, {Michtchenko}, {Miglio}, {Miguel}, {Milligan}, {Mirouh}, {Mitchell}, {Moedas}, {Molendini}, {Moln{\'a}r}, {Mombarg}, {Montalban}, {Montalto}, {Monteiro}, {Montoro S{\'a}nchez}, {Morales}, {Morales-Calderon}, {Morbidelli}, {Mordasini}, {Moreau}, {Morel}, {Morello}, {Morin}, {Mortier}, {Mosser}, {Mourard}, {Mousis}, {Moutou}, {Mowlavi}, {Moya}, {Muehlmann}, {Muirhead}, {Munari}, {Musella}, {Mustill}, {Nardetto}, {Nardiello}, {Narita}, {Nascimbeni}, {Nash}, {Neiner}, {Nelson}, {Nettelmann}, {Nicolini}, {Nielsen}, {Niemi}, {Noack}, {Noels-Grotsch}, {Noll}, {Norazman}, {Norton}, {Nsamba}, {Ofir}, {Ogilvie}, {Olander}, {Olivetto}, {Olofsson}, {Ong}, {Ortolani},
  {Oshagh}, {Ottacher}, {Ottensamer}, {Ouazzani}, {Paardekooper}, {Pace}, {Pajas}, {Palacios}, {Palandri}, {Palle}, {Paproth}, {Parro}, {Parviainen}, {Granado}, {Passegger}, {Pastor-Morales}, {P{\"a}tzold}, {Gade Pedersen}, {Pena Hidalgo}, {Pepe}, {Pereira}, {Persson}, {Pertenais}, {Peter}, {Petit}, {Petit}, {Pezzuto}, {Pichierri}, {Pietrinferni}, {Pinheiro}, {Pinsonneault}, {Plachy}, {Plasson}, {Plez}, {Poppenhaeger}, {Poretti}, {Portaluri}, {Portell}, {Frederico Porto de Mello}, {Poyatos}, {Pozuelos}, {Prada Moroni}, {Pricopi}, {Prisinzano}, {Quade}, {Quirrenbach}, {Rabanal Reina}, {Rabello Soares}, {Raimondo}, {Rainer}, {Ram{\'o}n Rod{\'o}n}, {Ram{\'o}n-Ballesta}, {Ramos Zapata}, {R{\"a}tz}, {Rauterberg}, {Redman}, {Redmer}, {Reese}, {Regibo}, {Reiners}, {Reinhold}, {Renie}, {Ribas}, {Ribeiro}, {Pereira Ricciardi}, {Rice}, {Richard}, {Riello}, {Rieutord}, {Ripepi}, {Rixon}, {Rockstein}, {Ram{\'o}n Rod{\'o}n Ortiz}, {Rodr{\'\i}guez}, {Rodr{\'\i}guez Amor}, {Rodr{\'\i}guez D{\'\i}az}, {Rodriguez Garcia},
  {Rodriguez-Gomez}, {Roehlly}, {Roig}, {Rojas-Ayala}, {Rolf}, {Lysgaard R{\o}rsted}, {Rosado}, {Rosotti}, {Roth}, {Roth}, {Rousseau}, {Roxburgh}, {Roy}, {Royer}, {Ruane}, {Rufini Mastropasqua}, {Ruiz de Galarreta}, {Russi}, {Saar}, {Saillenfest}, {Salaris}, {Salmon}, {Saltas}, {Samadi}, {Samadi}, {Samra}, {Sanches da Silva}, {Andr{\'e}s S{\'a}nchez Carrasco}, {Santerne}, {Santiago P{\'e}}, {Santoli}, {Santos}, {Sanz Mesa}, {Sarro}, {Scandariato}, {Sch{\"a}fer}, {Schlafly}, {Schmider}, {Schneider}, {Schou}, {Schunker}, {J{\"o}rg Schwarzkopf}, {Serenelli}, {Seynaeve}, {Shan}, {Shapiro}, {Shipman}, {Sicilia}, {Sierra sanmartin}, {Sigot}, {Silliman}, {Silvotti}, {Simon}, {Simoyama Napoli}, {Skarka}, {Smalley}, {Smiljanic}, {Smit}, {Smith}, {Smith}, {Snellen}, {S{\'o}dor}, {Sohl}, {Solanki}, {Sortino}, {Sousa}, {Southworth}, {Souto}, {Sozzetti}, {Stamatellos}, {Stassun}, {Steller}, {Stello}, {Stelzer}, {Stiebeler}, {Stokholm}, {Storelvmo}, {Strassmeier}, {Str{\o}m}, {Strugarek}, {Sulis}, {vanda}, {Szabados},
  {Szab{\'o}}, {Szab{\'o}}, {Szuszkiewicz}, {Talens}, {Teti}, {Theisen}, {Th{\'e}venin}, {Thoul}, {Tiphene}, {Titz-Weider}, {Tkachenko}, {Tomecki}, {Tonfat}, {Tosi}, {Trampedach}, {Traven}, {Triaud}, {Tr{\o}nnes}, {Tsantaki}, {Tschentscher}, {Turin}, {Tvaruzka}, {Ulmer}, {Ulmer-Moll}, {Ulusoy}, {Umbriaco}, {Valencia}, {Valentini}, {Valio}, {Valverde Guijarro}, {Van Eylen}, {Van Grootel}, {van Kempen}, {Van Reeth}, {Van Zelst}, {Vandenbussche}, {Vasiliou}, {Vasilyev}, {Vaz de Mascarenhas}, {Vazan}, {Vela Nunez}, {Nunes Velloso}, {Ventura}, {Ventura}, {Venturini}, {Trallero}, {Veras}, {Verdugo}, {Verma}, {Vibert}, {Vicanek Martinez}, {Vida}, {Vigan}, {Villacorta}, {Villaver}, {Villaverde Aparicio}, {Viotto}, {Vorobyov}, {Vorontsov}, {Wagner}, {Walloschek}, {Walton}, {Walton}, {Wang}, {Waters}, {Watson}, {Wedemeyer}, {Weeks}, {Weingrill}, {Weiss}, {Wendler}, {West}, {Westerdorff}, {Westphal}, {Wheatley}, {White}, {Whittaker}, {Wickhusen}, {Wilson}, {Windsor}, {Winter}, {Lykke Winther}, {Winton}, {Witteck},
  {Witzke}, {Woitke}, {Wolter}, {Wuchterl}, {Wyatt}, {Yang}, {Yu}, {Zanmar Sanchez}, {Rosa Zapatero Osorio}, {Zechmeister}, {Zhou}, {Ziemke}, \& {Zwintz}}]{Rauer2024}
{Rauer}, H., {Aerts}, C., {Cabrera}, J., {et~al.} 2024, arXiv e-prints, arXiv:2406.05447

\bibitem[{{Ricker} {et~al.}(2015){Ricker}, {Winn}, {Vanderspek}, {Latham}, {Bakos}, {Bean}, {Berta-Thompson}, {Brown}, {Buchhave}, {Butler}, {Butler}, {Chaplin}, {Charbonneau}, {Christensen-Dalsgaard}, {Clampin}, {Deming}, {Doty}, {De Lee}, {Dressing}, {Dunham}, {Endl}, {Fressin}, {Ge}, {Henning}, {Holman}, {Howard}, {Ida}, {Jenkins}, {Jernigan}, {Johnson}, {Kaltenegger}, {Kawai}, {Kjeldsen}, {Laughlin}, {Levine}, {Lin}, {Lissauer}, {MacQueen}, {Marcy}, {McCullough}, {Morton}, {Narita}, {Paegert}, {Palle}, {Pepe}, {Pepper}, {Quirrenbach}, {Rinehart}, {Sasselov}, {Sato}, {Seager}, {Sozzetti}, {Stassun}, {Sullivan}, {Szentgyorgyi}, {Torres}, {Udry}, \& {Villasenor}}]{Ricker2015}
{Ricker}, G.~R., {Winn}, J.~N., {Vanderspek}, R., {et~al.} 2015, Journal of Astronomical Telescopes, Instruments, and Systems, 1, 014003

\bibitem[{{Santos} {et~al.}(2021){Santos}, {Breton}, {Mathur}, \& {Garc{\'\i}a}}]{Santos2021}
{Santos}, {\^A}.~R.~G., {Breton}, S.~N., {Mathur}, S., \& {Garc{\'\i}a}, R.~A. 2021, \apjs, 255, 17

\bibitem[{{Santos} {et~al.}(2019){Santos}, {Garc{\'\i}a}, {Mathur}, {Bugnet}, {van Saders}, {Metcalfe}, {Simonian}, \& {Pinsonneault}}]{Santos2019}
{Santos}, {\^A}.~R.~G., {Garc{\'\i}a}, R.~A., {Mathur}, S., {et~al.} 2019, \apjs, 244, 21

\bibitem[{Seager(2010)}]{Seager2010}
Seager, S. 2010, Exoplanet atmospheres: physical processes, Princeton series in astrophysics (Princeton, N.J: Princeton University Press), oCLC: ocn466343197

\bibitem[{{Shporer}(2017)}]{Shporer2017}
{Shporer}, A. 2017, \pasp, 129, 072001

\bibitem[{{Skrutskie} {et~al.}(2006){Skrutskie}, {Cutri}, {Stiening}, {Weinberg}, {Schneider}, {Carpenter}, {Beichman}, {Capps}, {Chester}, {Elias}, {Huchra}, {Liebert}, {Lonsdale}, {Monet}, {Price}, {Seitzer}, {Jarrett}, {Kirkpatrick}, {Gizis}, {Howard}, {Evans}, {Fowler}, {Fullmer}, {Hurt}, {Light}, {Kopan}, {Marsh}, {McCallon}, {Tam}, {Van Dyk}, \& {Wheelock}}]{Skrutskie2006}
{Skrutskie}, M.~F., {Cutri}, R.~M., {Stiening}, R., {et~al.} 2006, \aj, 131, 1163

\bibitem[{{Strugarek}(2018)}]{Strugarek2018}
{Strugarek}, A. 2018, in Handbook of Exoplanets, ed. H.~J. {Deeg} \& J.~A. {Belmonte}, 25

\bibitem[{{Torrence} \& {Compo}(1998)}]{Torrence1998}
{Torrence}, C. \& {Compo}, G.~P. 1998, Bulletin of the American Meteorological Society, 79, 61

\bibitem[{{Torres} {et~al.}(2017){Torres}, {Kane}, {Rowe}, {Batalha}, {Henze}, {Ciardi}, {Barclay}, {Borucki}, {Buchhave}, {Crepp}, {Everett}, {Horch}, {Howard}, {Howell}, {Isaacson}, {Jenkins}, {Latham}, {Petigura}, \& {Quintana}}]{2017Torres}
{Torres}, G., {Kane}, S.~R., {Rowe}, J.~F., {et~al.} 2017, \aj, 154, 264

\bibitem[{Virtanen {et~al.}(2020)Virtanen, Gommers, Oliphant, Haberland, Reddy, Cournapeau, Burovski, Peterson, Weckesser, Bright, {van der Walt}, Brett, Wilson, Millman, Mayorov, Nelson, Jones, Kern, Larson, Carey, Polat, Feng, Moore, {VanderPlas}, Laxalde, Perktold, Cimrman, Henriksen, Quintero, Harris, Archibald, Ribeiro, Pedregosa, {van Mulbregt}, \& {SciPy 1.0 Contributors}}]{2020SciPy-NMeth}
Virtanen, P., Gommers, R., Oliphant, T.~E., {et~al.} 2020, Nature Methods, 17, 261

\bibitem[{Wiecki {et~al.}(2023)Wiecki, Salvatier, Vieira, Kochurov, Patil, Osthege, Willard, \& Engels}]{pymc}
Wiecki, T., Salvatier, J., Vieira, R., {et~al.} 2023, pymc-devs/pymc: v5.4.0

\bibitem[{Winn(2014)}]{Winn2014}
Winn, J.~N. 2014, Transits and {Occultations}, arXiv:1001.2010 [astro-ph]

\bibitem[{{Woodard}(1984)}]{Woodard1984}
{Woodard}, M.~F. 1984, PhD thesis, University of California, San Diego.

\bibitem[{{Zahn}(1975)}]{1975Zahn}
{Zahn}, J.~P. 1975, \aap, 41, 329

\end{thebibliography}

\begin{appendix}

\section{Contaminants \label{Appendix:AppendixA}}

\subsection{Transiting exoplanet \label{App:transit}}
In Sect.~\ref{sec:comparison with the litterature}, we were able to remove 4 stars confirmed to harbour transiting exoplanets. For example, \citet{2016Morton} discovered Kepler-685b, a gas giant exoplanet orbiting around KIC 3351888 with an orbital period of 1.62 days. 

\begin{figure}[htbp]
    \centering
    \includegraphics[width=0.45\textwidth]{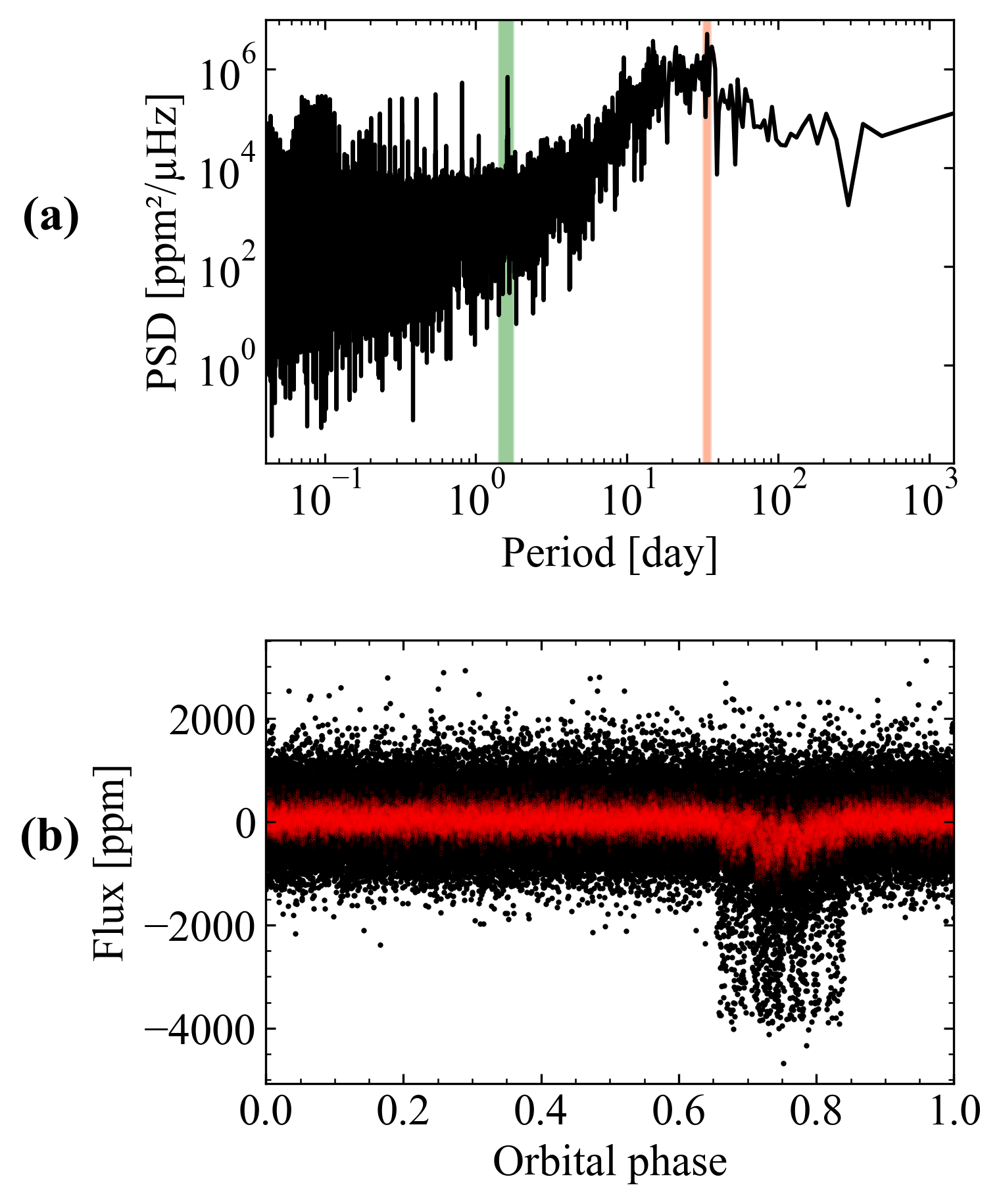}
    \caption{(a) PSD of KIC 3351888. The \prot, shown in the orange area, was measured by \citet{Santos2021}. The green area shows the high-amplitude peak signal detected in this work, at around 1.62 days. (b) Phase-folding analysis on the light curve of KIC 3351888. The red curve is a smoothing of the black curve, obtained with a triangular filter.}
    \label{fig:phase_fold_transit}
\end{figure}

Panel (a) of Fig. \ref{fig:phase_fold_transit} shows the PSD of KIC 3351888 where a high-amplitude peak signal at $1.62$ days can be observed. In panel (b) of Fig \ref{fig:phase_fold_transit}, the phase-folding curve demonstrates a flux variation analogous to that observed during a transit. The signal detected in this work for KIC 3351888 therefore corresponds to the planet Kepler-685b. Because it has already been detected, this candidate is not part of our final sample.

\subsection{Phase-folding\label{App:phasefold}}
By phase-folding the light curve, we can examine the temporal stability of a signal corresponding to a given period. To refine our sample, we performed a visual inspection of the 283 stars obtained after Sects.~\ref{sec:psd} and \ref{sec:comparison with the litterature}. In these sections, we searched for high-amplitude peaks in the PSDs and compared our results with existing catalogues. In Sect.~\ref{sec:phasefold,wps}, we performed a phase-folding analysis followed by a visual inspection.

\begin{figure}[htbp]
    \raggedright
    \includegraphics[width=0.45\textwidth]{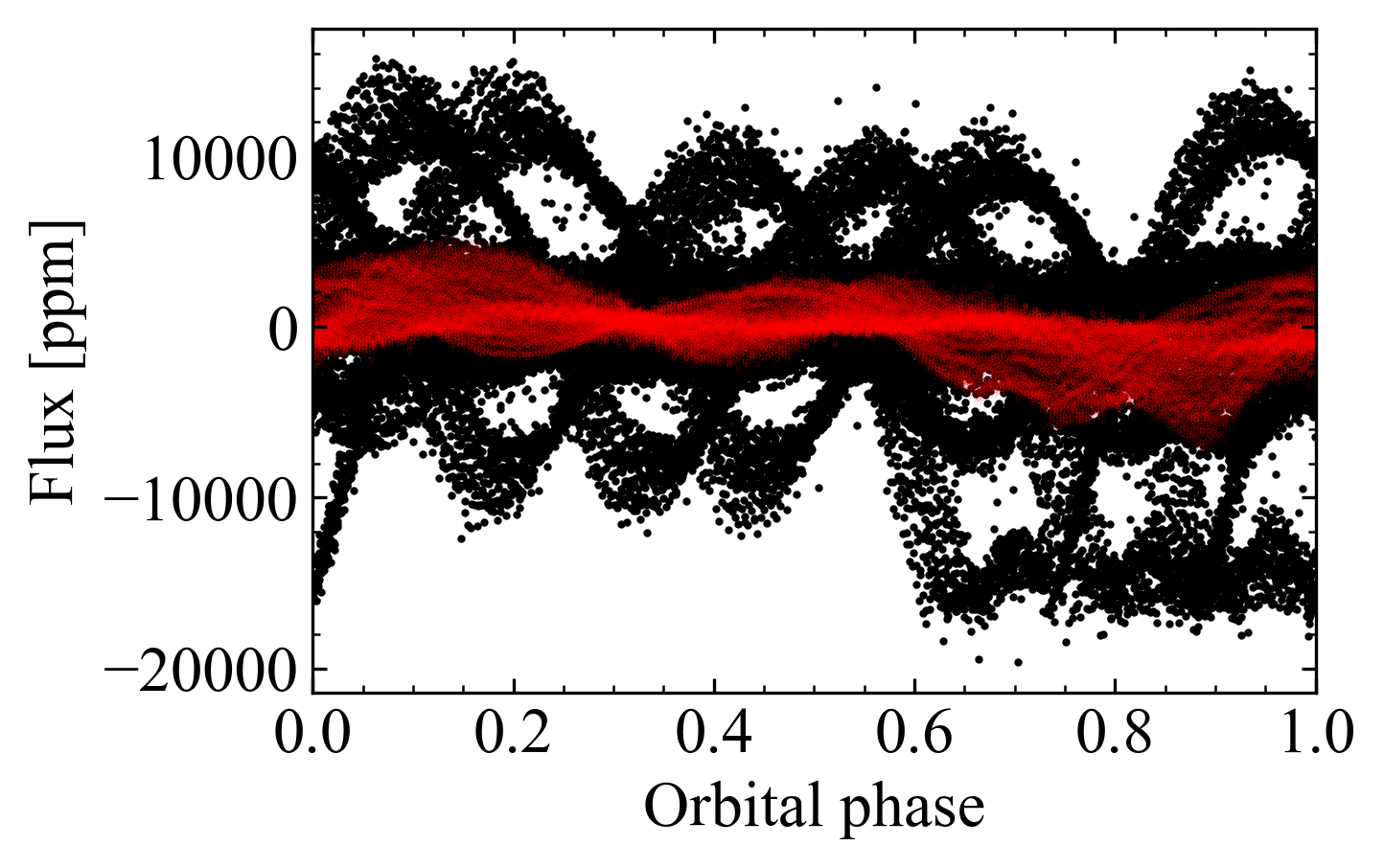}
    \caption{Phase-folding analysis on the light curve of KIC~11154093. The red curve is a smoothing of the black curve, obtained with a triangular filter.}
    \label{fig:phase_fold_app}
\end{figure}

Fig.~\ref{fig:phase_fold_app} shows the phase-folded light curve of KIC 11154093, which exhibits a complex pattern with multiple distinct flux variations across the folding period. Visually, the observed pattern deviates significantly from the expected behaviour of a signal (see Fig.~\ref{fig:composantes} and panel (c) of Fig.~\ref{fig:KIC_example}). This complex behaviour might be linked to the presence of multiple stars close to this target, such as seen when looking at the 2MASS image (see Appendix~\ref{App:KASOC}) of the target in Fig.~\ref{fig:2MASSKIC}. As a result, this target was excluded from our sample.

\begin{figure}
    \centering
    \includegraphics[width=0.7\linewidth]{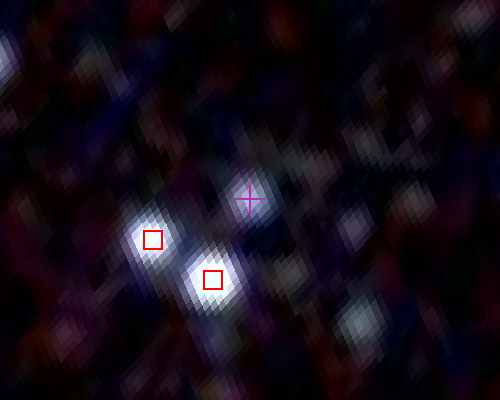}
    \caption{Image from the KASOC website of KIC 11154093 using 2MASS \citep{Skrutskie2006}. The purple cross marks KIC 11154093.}
    \label{fig:2MASSKIC}
\end{figure}

\subsection{Wavelet analysis\label{App:wavelet}}
\begin{figure}[htbp]
    \includegraphics[width=0.5\textwidth]{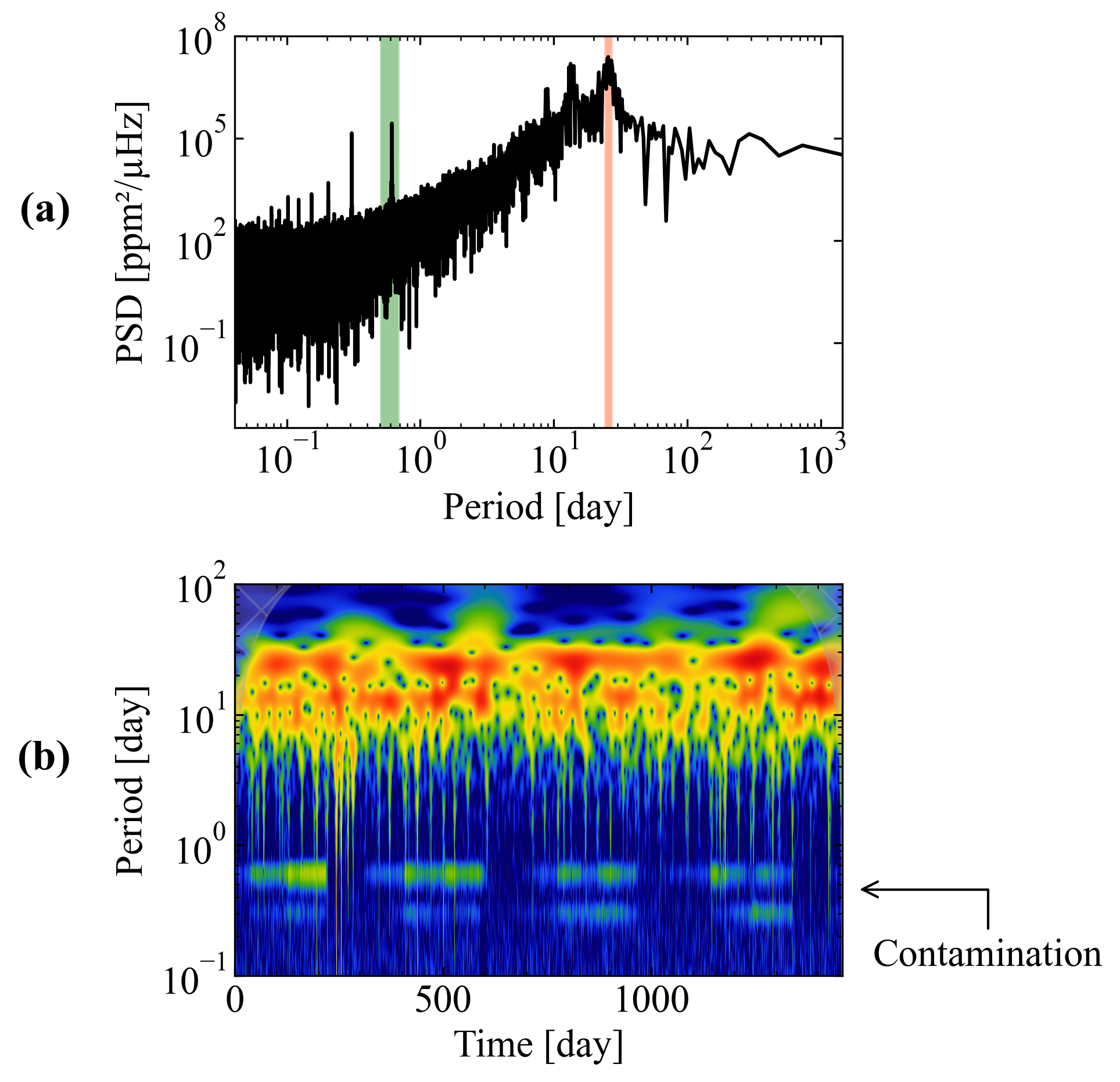}
    \caption{Rotational analysis of KIC 3233612. (a) PSD of the star. The stellar rotation period is shown in the orange area, while the green area shows the high-amplitude peak at around 0.6 days. (b) Period-time analysis of the stellar light curve. High-amplitude periodicity is highlighted by orange and red, whereas yellow and blue denote low-power amplitude. The yellow discontinuous structure appearing around 0.6 days results from background contamination. The shaded area represents the cone of influence of the reliable periodicities.}
    \label{fig:Wavelet_analysis}
\end{figure}

In Sect.~\ref{sec:phasefold,wps}, we computed the WPS of each of the 245 stellar light curves, and we only kept stars that present stable modulations along time, as shown in panel (d) of Fig. \ref{fig:KIC_example}. \\
Figure \ref{fig:Wavelet_analysis} presents the rotational analysis of KIC 2852669, a K-type star included in our sample of 245 stars. The top panel displays the PSD while the bottom panel shows the normalised time-averaged projection of the wavelet transform. 

The different analyses all point to an orbital period of around 0.6 days (panel (a), Fig.~\ref{fig:Wavelet_analysis}). However, the signal shown on panel (b) of Fig.~\ref{fig:Wavelet_analysis} does not show stable modulations along time. We recall that the focal plane of \kep is rotated after each quarter, so the photometric apertures used to monitor a given target have varying levels of background contamination. The periodic signal found in this study is therefore unlikely to be caused by a non-transiting object, and KIC 3233612 was excluded from our sample. 

\newpage
\subsection{Field contaminants \label{App:KASOC}}

In Sect.~\ref{sec:phasefold,wps}, targets with a nearby bright object in their aperture are removed from the sample.
We first looked for potential contaminants using the \textit{Gaia} catalogue \citep{2023GaiaCollaboration} and found 70 \textit{Gaia} objects near 38 of our targets. We excluded 11 targets for which the potential contaminants are located within a few arcseconds. The 27 remaining stars are located at least 10'' away from the potential contaminant, making significant contaminations problems unlikely. The 10 closest contaminants, associated with these 27 retained targets, are listed in Table~\ref{tb:Gaia_discardedKICs}.

\renewcommand{\arraystretch}{0.7}
\begin{table}[htbp]\small
    \centering
    \caption{Candidates with a contaminant located at least 10'' away and less than 40'' away from the considered star.}
    \resizebox{0.49\textwidth}{!}{
    \begin{tabular}{lccc}
    \hline \hline \\
    KIC & \textit{Gaia} ID & Contaminant \textit{Gaia} ID & distance \\ \\ 
    \hline \\
    3222610  & 2099534981005352576 & 2099535015365093120 & 17.88 \\ \\
    5731359  & 2073820977455843200 & 2073821183614272256 & 16.51 \\ \\
    5938531  & 2104466423796556160 & 2104466423796548992 & 19.02 \\ \\
    6370174  & 2077612883814892160 & 2077614361283648128 & 20.54 \\ \\
    7511885  & 2105555322564080384 & 2105555322564081664 & 18.85 \\ \\
    7882472  & 2103046030870052224 & 2103046030870053760 & 18.29 \\ \\
    10027247 & 2086316205543022336 & 2086316205543023104 & 14.29 \\ \\
    10291936 & 2086348091380882688 & 2086348160100359808 & 13.29 \\ \\
    11408355 & 2134998079291667968 & 2135001034229167232 & 11.04 \\ \\
    11605209 & 2132673883508778880 & 2132673883508775168 & 18.45 \\ \\
    ...      & ...                 &     ...             & ...  \\ \\
    \bottomrule
    \end{tabular}}
    \tablefoot{Only the 10 closest contaminants are shown in this table. \textit{Gaia} ID is the correponding \textit{Gaia} ID of the \kep star. The distance, measured as the angular separation between the star and the contaminant, is expressed in arcseconds.}
    \label{tb:Gaia_discardedKICs}
\end{table}

\begin{figure}[h!]
    \centering
    \includegraphics[width=0.4\textwidth]{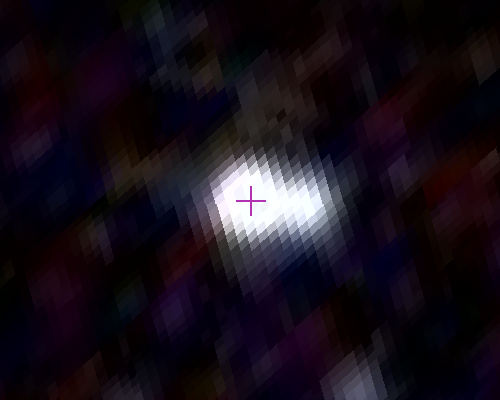}
    \caption{Image from the KASOC website of KIC 12505054 using 2MASS \citep{Skrutskie2006}. The purple cross marks KIC 12505054.}
    \label{fig:KASOC_check}
\end{figure}

We then carried out a visual inspection using the 2MASS images \citep{Skrutskie2006}, which can be accessed through the KASOC database \citep{2010Kjeldsen}. Similarly to what was done e.g. in \citet{Noraz2022}, we excluded candidates if the star is within the halo of a brighter object.

Figure \ref{fig:KASOC_check} shows KIC 12505054, a star located within the halo of another bright object not listed in the KASOC database. As a result, this star was excluded from our final sample. Indeed, a nearby bright object is likely to contaminate the signal obtained for this star.

\onecolumn

\section{Summary of photometric fit results \label{App:results_fit}}
\begin{figure*}[h!]
    \centering
    \includegraphics[width=0.9\textwidth]{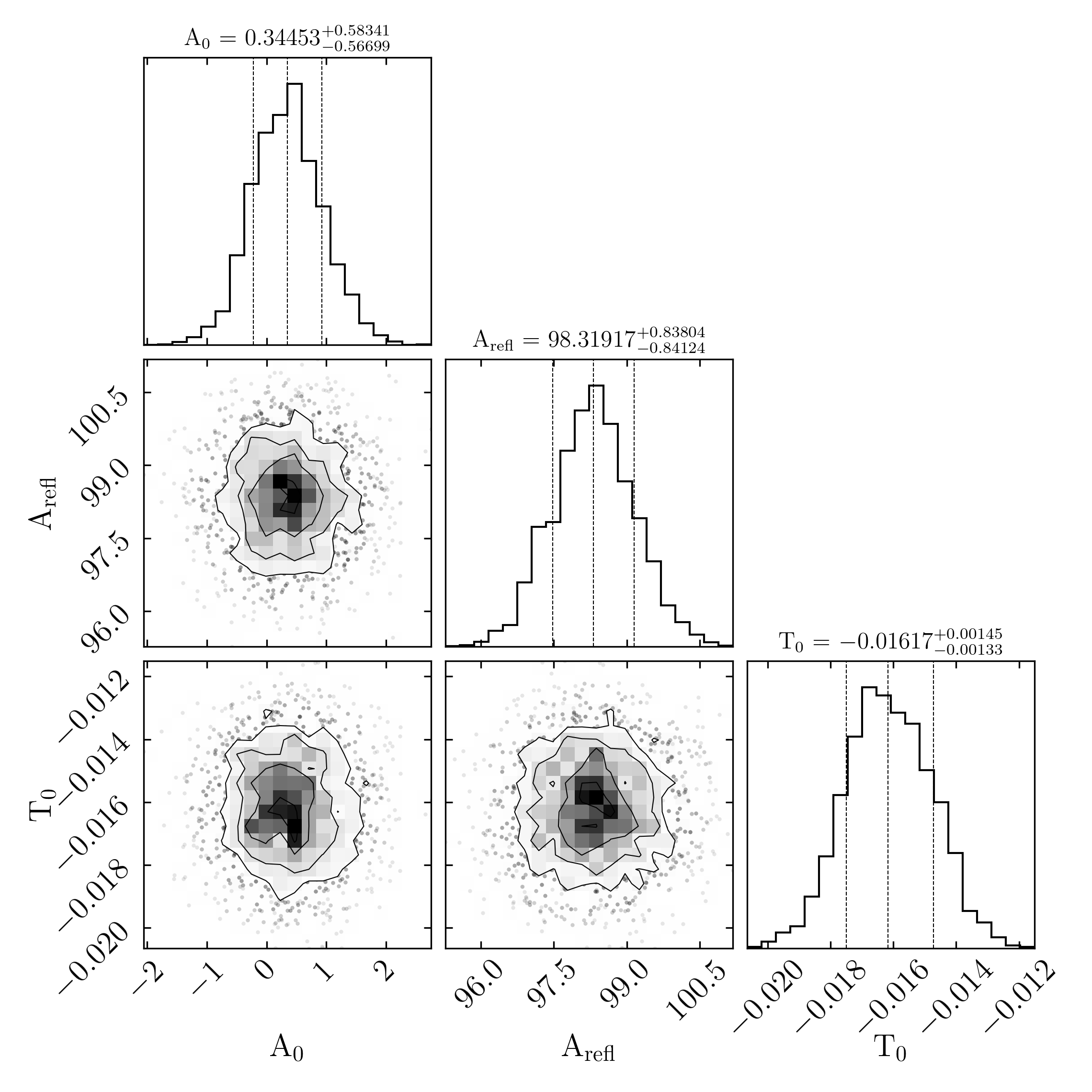}
    \caption{Corner plot for the fit of KIC~4930913 showing the posterior distributions of the fitted parameters A$_0$, A$_{\rm{atm}}$ and T$_0$. The values and error bars represent the median and 1$\sigma$ uncertainty based on the posterior distributions.}
    \label{fig:corner_kic_4930913}
\end{figure*}

In Sect.~\ref{sec:fit of the photometric data} we computed the fit of the photometric modulation data and retrieved the theoretical values of $A_{\rm atm}$. By doing so, we were able to deduce the parameter $\alpha_{\mathrm{refl}} \sin i \left(R_{\rm p}/R_{\mathrm{J}}\right)^2$, thanks to Eq.~(\ref{eq:atm}). These resulting parameters are presented in Table~\ref{tab:Aatm fits} for the 15 first targets. Fig.~\ref{fig:corner_kic_4930913} shows the corner plot of the posterior probability distribution resulting from the fit of KIC~4930913, a star introduced in Sect.~\ref{sec:fit of the photometric data} to illustrate the fitting of light curves exhibiting approximately sinusoidal modulation. The diagonal panels display the marginalised one dimensional distribution for each parameter: the baseline amplitude $A_0$, the atmospherical amplitude $A_{\rm atm}$ and the reference time $T_0$. The off-diagonal panels present the two-dimensional joint posterior distributions, with contour levels corresponding to the $1\sigma$, $2\sigma$, and $3\sigma$ credible regions. The results indicate that all parameters are well constrained. In particular, $T_0$ shows a narrow distribution, implying high precision in the timing estimate. The distribution of $A_0$ is broader and slightly asymmetric, reflecting larger uncertainties. No significant correlations are observed between the parameters, suggesting they are independently constrained by the data.
We also calculated with Eq.~(\ref{eq:imax}) the maximum inclination possible for each system, assuming they all follow a circular orbit. For two stars (KIC~3124224 and KIC~10396339), we do not provide $i_{\rm max}$, because, using $R_\star$ and $M_\star$ from the reference catalogues adopted, we obtain $a < R_\star$, which is obviously unphysical. Nevertheless, considering that $R_\star$ and $M_\star$ might be incorrect, we decided to keep these two stars in our sample.

\begin{figure}[h!]
    \centering
    \includegraphics[width=0.9\textwidth]{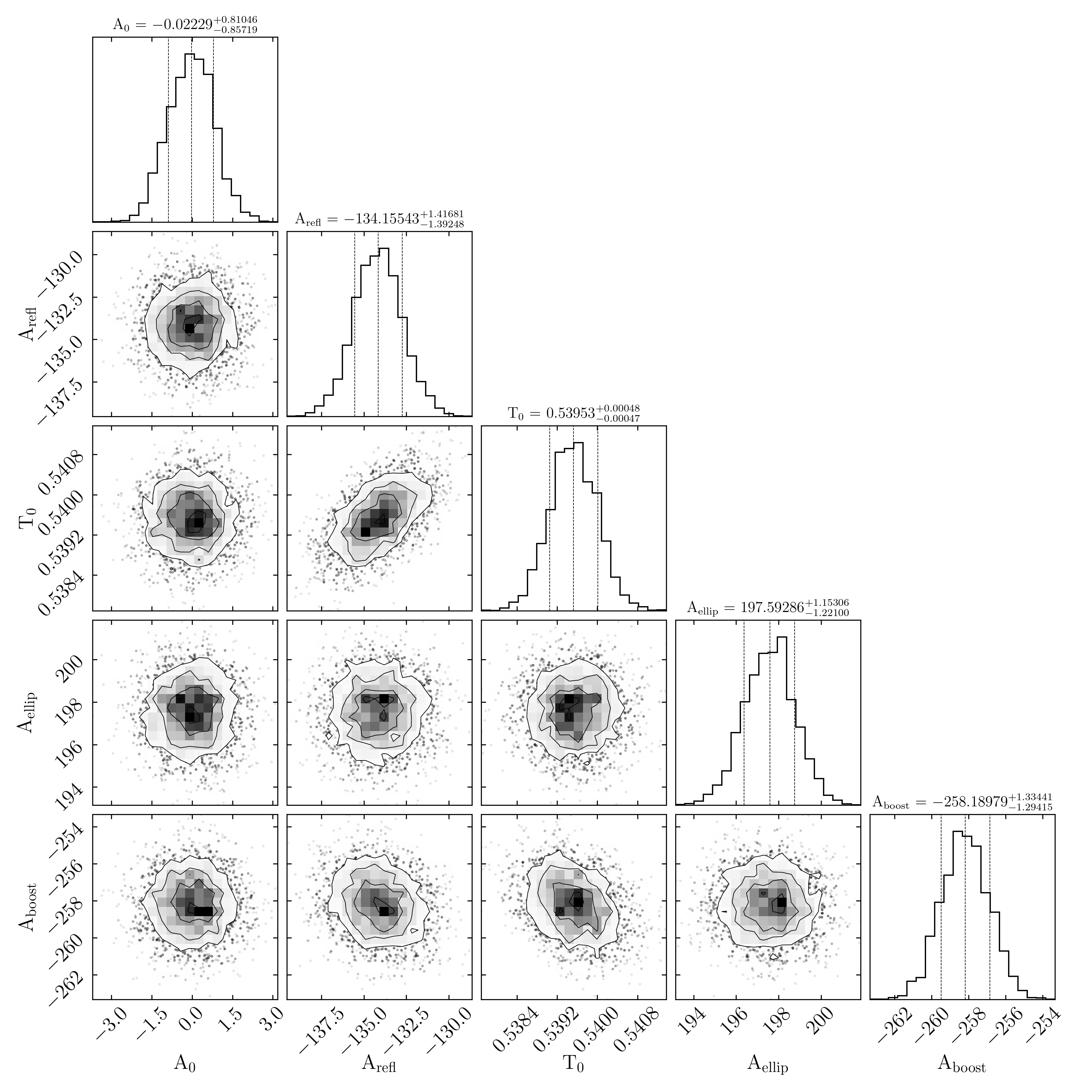}
    \caption{Corner plot for the fit of KIC~4264634 showing the posterior distributions of the fitted parameters A$_0$, A$_{\rm{atm}}$, A$_{\rm{ellip}}$, A$_{\rm{boost}}$ and T$_0$. The values and error bars represent the median and 1$\sigma$ uncertainty based on the posterior distributions.}
    \label{fig:corner_kic_4264634}
\end{figure}

When the modulation of the phase-folded light curve was more complex, we performed a fit considering the three components of the signal. The corner plot in Fig.~\ref{fig:corner_kic_4264634} summarises the posterior probability distributions resulting for KIC~4264634, the example star chosen in Sect.~3 to illustrate the fit which accounts for the three components of the signal. As in Fig.~\ref{fig:corner_kic_4930913}, the diagonal panels show the marginalised one-dimensional distributions for each fitted parameter: the baseline amplitude, $A_0$, the atmospheric amplitude, $A_{\rm atm}$, the reference time, $T_0$, the ellipsoidal amplitude, $A_{\rm ellip}$, and the Doppler boosting amplitude, $A_{\rm boost}$. The off-diagonal panels illustrate the two-dimensional joint posterior distributions, with contours corresponding to the $1\sigma$, $2\sigma$, and $3\sigma$ credible intervals. All parameters are reasonably well constrained, with the amplitude parameters displaying broader distributions, especially $A_0$ and $A_{\rm atm}$, which exhibit mild asymmetries. The joint distributions suggest weak correlations between some of the amplitude parameters, particularly between $A_{\rm atm}$ and $A_{\rm ellip}$. Overall, the parameters remain largely uncorrelated. This supports the robustness of the fit and the ability of the model to independently constrain each physical contribution. The resulting parameters for all of the nine targets studied here are presented in Table~\ref{tab:Agrav fits}. In both Tables~\ref{tab:Aatm fits} and \ref{tab:Agrav fits}, the reported values of $A_{\rm atm}, A_{\rm beam}$ and $A_{\rm ellip}$ correspond to their absolute values. Thanks to Eq.~(\ref{eq:ellip}) and (\ref{eq:boost}) we were also able to determine the parameters $\left(M_{\rm p} \sin^{2}i/M_{\mathrm{J}}\right)$ and $\left(M_{\rm p} \sin i/M_{\mathrm{J}}\right)$. With such calculations, we can in principle retrieve the value of inclination of the non-transiting object $i$. This was possible for three of our systems (KIC~11550689, 5697777, and 4264634). For three other stars, the uncertainties on our calculated parameters are too large to infer on the value $i$. For the last three stars, the uncertainties are not sufficiently high enough to explain the obtained inconsistent results. We are therefore led to believe that these stars are out of the validity range of the model used in this article and would probably require more complex models \citep[see][]{2019Penoyre}.
\FloatBarrier

\renewcommand{\arraystretch}{0.8}
\begin{table}[ht!]
    \caption{Photometric fit results of 15 out of the 77 candidates showing quasi-sinusoidal modulations}
    \centering
    \begin{tabular}{cccc}
    \hline \hline \\
    KIC & $A_{\rm atm}$ [ppm] & $\alpha_{\mathrm{refl}} \sin i \left(R_{\rm p}/R_{\mathrm{J}}\right)^2$ & $i_{\rm max}$ [\textdegree]\\ \\
    \hline\\
    2707504 & $185.91^{+0.63}_{-0.64}$ & $6.3026^{+0.0214}_{-0.0216}$ & 82 \\ \\
    3124224 & $72.51^{+0.74}_{-0.76}$ & $0.1416^{+0.0014}_{-0.0015}$  & - \\ \\
    3222610 & $133.07^{+0.96}_{-0.96}$ & $1.8074^{+0.0130}_{-0.0131}$ & 80 \\ \\
    3757251 & $233.86^{+2.48}_{-2.66}$ & $0.7963^{+0.0084}_{-0.0091}$ & 73 \\ \\
    3954961 & $218.78^{+1.24}_{-1.18}$ & $3.2079^{+0.0181}_{-0.0174}$ & 79 \\ \\
    4175216 & $279.84^{+0.80}_{-0.78}$ & $2.9049^{+0.0083}_{-0.0081}$ & 78 \\ \\
    4283417 & $190.32^{+1.83}_{-1.82}$ & $2.4066^{+0.0231}_{-0.0230}$ & 80 \\ \\
    4641549 & $92.29^{+2.31}_{-2.34}$ & $1.9074^{+0.0477}_{-0.0484}$ & 79 \\ \\
    4859400 & $40.74^{+0.75}_{-0.69}$ & $0.7340^{+0.0135}_{-0.0125}$ & 67 \\ \\
    4862155 & $65.28^{+0.73}_{-0.72}$ & $1.4593^{+0.0163}_{-0.0160}$ & 72 \\ \\
    4916039 & $93.12^{+1.87}_{-1.84}$ & $0.4643^{+0.0093}_{-0.0092}$ & 71 \\ \\
    4930913 & $98.32^{+0.84}_{-0.84}$ & $1.2077^{+0.0103}_{-0.0103}$ & 73 \\ \\
    4945547 & $60.85^{+0.97}_{-1.00}$ & $0.4127^{+0.0066}_{-0.0068}$ & 74 \\ \\
    4996173 & $39.76^{+0.14}_{-0.14}$ & $0.3498^{+0.0012}_{-0.0012}$ & 49 \\ \\
    5006208 & $52.85^{+0.71}_{-0.62}$ & $2.6370^{+0.0356}_{-0.0312}$ & 75 \\ \\
    \vdots & \vdots & \vdots & \vdots \\ \\
    \hline
    \end{tabular}
    \label{tab:Aatm fits}
\end{table}

\begin{table}[ht!]
    \caption{Photometric fit results for the nine candidates showing complex modulations}
    \centering
    \begin{tabular}{cccccccc}
    \hline \hline \\
    KIC & $A_{\rm atm}$ [ppm] & $\alpha_{\mathrm{refl}} \sin i \left(R_{\rm p}/R_{\mathrm{J}}\right)^2$ & $A_{\rm boost}$ [ppm] & $A_{\rm ellip}$ [ppm]& $\left(M_{\rm p} \sin i/M_{\mathrm{J}}\right)$ & $\left(M_{\rm p} \sin^{2}i/M_{\mathrm{J}}\right)$ & $i_{\rm max}$ [\textdegree] \\ \\
    \hline\\
    4264634 & $134.15^{+1.41}_{-1.38}$ & $1.3686^{+0.0144}_{-0.0141}$ & $258.19^{+1.34}_{-1.29}$ & $197.59^{+1.16}_{-1.22}$ & $12.97^{+0.07}_{-0.06}$ & $8.02^{+0.05}_{-0.05}$ & 78 \\ \\
    4373708 & $301.70^{+1.31}_{-1.41}$ & $1.4014^{+0.0061}_{-0.0065}$ & $9.12^{+1.84}_{-1.79}$ & $163.97^{+1.33}_{-1.24}$ & $0.45^{+0.09}_{-0.09}$ & $1.44^{+0.01}_{-0.01}$ & 69 \\ \\
    5440365 & $0.73^{+0.85}_{-0.86}$ & $0.0156^{+0.0181}_{-0.0184}$ & $0.30^{+0.84}_{-0.83}$ & $31.52^{+0.85}_{-0.85}$ & $0.03^{+0.09}_{-0.09}$ & $1.81^{+0.05}_{-0.05}$ & 76 \\ \\
    5622796 & $2.42^{+0.79}_{-0.73}$ & $0.1068^{+0.0348}_{-0.0320}$ & $2.32^{+0.79}_{-0.81}$ & $45.29^{+0.82}_{-0.75}$ & $0.28^{+0.09}_{-0.10}$ & $9.33^{+0.17}_{-0.15}$ & 81 \\ \\
    5697777 & $7.42^{+0.62}_{-0.60}$ & $0.0766^{+0.0064}_{-0.0062}$ & $78.80^{+0.47}_{-0.44}$ & $41.29^{+0.47}_{-0.42}$ & $9.89^{+0.06}_{-0.06}$ & $0.07^{+0.00}_{-0.00}$ & 16 \\ \\
    8844761 & $0.57^{+0.60}_{-0.55}$ & $0.0049^{+0.0052}_{-0.0048}$ & $0.30^{+0.55}_{-0.57}$ & $25.94^{+0.56}_{-0.57}$ & $0.01^{+0.03}_{-0.03}$ & $0.89^{+0.02}_{-0.02}$ & 78 \\ \\
    11550689 & $2.09^{+1.44}_{-1.49}$ & $0.0053^{+0.0036}_{-0.0037}$ & $88.35^{+1.21}_{-1.22}$ & $62.71^{+1.27}_{-1.13}$ & $3.31^{+0.05}_{-0.05}$ & $0.34^{+0.01}_{-0.01}$ & 66 \\ \\
    11668095 & $3.17^{+2.39}_{-2.44}$ & $0.0823^{+0.0620}_{-0.0635}$ & $2.26^{+2.46}_{-2.48}$ & $179.92^{+2.35}_{-2.34}$ & $0.14^{+0.15}_{-0.15}$ & $39.18^{+0.51}_{-0.51}$ & 83 \\ \\
    11702835 & $6.82^{+1.40}_{-1.51}$ & $0.0355^{+0.0073}_{-0.0079}$ & $35.77^{+1.49}_{-1.50}$ & $156.55^{+1.43}_{-1.56}$ & $1.74^{+0.07}_{-0.07}$ & $2.93^{+0.03}_{-0.03}$ & 74 \\ \\
    \hline
    \end{tabular}
    \label{tab:Agrav fits}
\end{table}

\section{Final sample \label{App:table_all}}

\renewcommand{\arraystretch}{0.8}
\addtolength{\tabcolsep}{-0.2em}
\begin{table}[ht!]
    \centering
    \caption{Global parameters of 16 out of the 88 candidates likely to host non-transiting companions. \label{tab:all_sample}}
    \begin{tabular}{ccccccccccc}
    \hline \hline \\
     & $P_{\rm orb}$ & semi-major axis & $A_{\rm periodogram}$ & $P_{\rm rot}$ & $T_{\rm eff}$  & Mass & Kpmag & log g & [Fe/H] & flag\\ \\
    KIC & [days] & [AU] & [ppm] & [days] & [K] & [$\rm M_\odot$] & [mag] & [dex] & [dex] & \\ \\
    \hline\\
    2707504 & $1.83354^{+0.00110}_{-0.00112}$ & $0.02721^{+0.00019}_{-0.00020}$ & $202^{+59}_{-39}$ & $18.3^{+1.9}_{-1.9}$ & $5135^{+179}_{-179}$ & $0.80^{+0.09}_{-0.07}$ & 13.68 & $4.56^{+0.05}_{-0.07}$ & $-0.06^{+0.30}_{-0.30}$ & 0, 0, 0 \\ \\
    3124224 & $0.16550^{+0.00001}_{-0.00001}$ & $0.01037^{+0.00001}_{-0.00001}$ & $73^{+22}_{-13}$ & $74.4^{+6.5}_{-6.5}$ & $5206^{+128}_{-122}$ & $1.36^{+0.04}_{-0.06}$ & 13.91 & $3.77^{+0.02}_{-0.03}$ & $0.28^{+0.16}_{-0.15}$ & 0, 0, 0 \\ \\
    3222610 & $1.01284^{+0.00038}_{-0.00028}$ & $0.01722^{+0.00009}_{-0.00007}$ & $144^{+90}_{-43}$ & $4.8^{+0.3}_{-0.3}$ & $4260^{+47}_{-70}$ & $0.66^{+0.02}_{-0.02}$ & 14.42 & $4.62^{+0.03}_{-0.02}$ & $0.06^{+0.12}_{-0.12}$ & 0, 0, 0 \\ \\
    3757251 & $0.44853^{+0.00008}_{-0.00009}$ & $0.00862^{+0.00003}_{-0.00003}$ & $228^{+46}_{-36}$ & $15.8^{+1.3}_{-1.3}$ & $3624^{+82}_{-86}$ & $0.42^{+0.01}_{-0.02}$ & 15.77 & $4.71^{+0.01}_{-0.01}$ & $0.05^{+0.12}_{-0.13}$ & 0, 0, 0 \\ \\
    3954961 & $1.04475^{+0.00023}_{-0.00021}$ & $0.01789^{+0.00007}_{-0.00006}$ & $238^{+74}_{-47}$ & $13.1^{+1.0}_{-1.0}$ & $4413^{+70}_{-61}$ & $0.70^{+0.03}_{-0.02}$ & 14.84 & $4.59^{+0.02}_{-0.01}$ & $0.20^{+0.11}_{-0.11}$ & 0, 0, 0 \\ \\
    4175216 & $0.93558^{+0.00035}_{-0.00038}$ & $0.01506^{+0.00008}_{-0.00008}$ & $272^{+41}_{-30}$ & $13.7^{+1.1}_{-1.1}$ & $4327^{+74}_{-69}$ & $0.52^{+0.03}_{-0.03}$ & 13.82 & $4.62^{+0.02}_{-0.02}$ & $0.01^{+0.11}_{-0.10}$ & 0, 0, 0 \\ \\
    4264634 & $0.88671^{+0.00038}_{-0.00051}$ & $0.01493^{+0.00008}_{-0.00010}$ & $316^{+134}_{-71}$ & $12.3^{+0.8}_{-0.8}$ & $3940^{+70}_{-94}$ & $0.56^{+0.02}_{-0.02}$ & 14.82 & $4.61^{+0.03}_{-0.02}$ & $0.16^{+0.12}_{-0.13}$ & 0, 0, 0 \\ \\
    4283417 & $0.99830^{+0.00061}_{-0.00057}$ & $0.01662^{+0.00012}_{-0.00011}$ & $185^{+41}_{-32}$ & $13.2^{+1.0}_{-1.0}$ & $4117^{+68}_{-68}$ & $0.61^{+0.02}_{-0.02}$ & 15.28 & $4.63^{+0.03}_{-0.03}$ & $0.06^{+0.12}_{-0.11}$ & 0, 0, 0 \\ \\
    4373708 & $0.46245^{+0.00013}_{-0.00014}$ & $0.01007^{+0.00004}_{-0.00004}$ & $293^{+43}_{-33}$ & $9.4^{+0.7}_{-0.7}$ & $4802^{+86}_{-80}$ & $0.64^{+0.04}_{-0.03}$ & 14.83 & $4.53^{+0.03}_{-0.03}$ & $0.09^{+0.12}_{-0.12}$ & 0, 0, 0 \\ \\
    4641549 & $1.18275^{+0.00034}_{-0.00038}$ & $0.02124^{+0.00009}_{-0.00010}$ & $101^{+63}_{-26}$ & $21.2^{+1.7}_{-1.7}$ & $5662^{+110}_{-103}$ & $0.91^{+0.05}_{-0.05}$ & 15.76 & $4.51^{+0.03}_{-0.04}$ & $-0.14^{+0.14}_{-0.16}$ & 0, 0, 0 \\ \\
    4859400 & $0.97366^{+0.00034}_{-0.00027}$ & $0.01984^{+0.00010}_{-0.00008}$ & $44^{+13}_{-8}$ & $21.5^{+2.7}_{-2.7}$ & $5875^{+134}_{-131}$ & $1.10^{+0.12}_{-0.08}$ & 14.00 & $4.02^{+0.06}_{-0.05}$ & $-0.06^{+0.15}_{-0.16}$ & 0, 0, 0 \\ \\
    4862155 & $1.04625^{+0.00044}_{-0.00042}$ & $0.02209^{+0.00012}_{-0.00012}$ & $71^{+18}_{-13}$ & $9.5^{+0.6}_{-0.6}$ & $6574^{+150}_{-141}$ & $1.31^{+0.06}_{-0.06}$ & 13.91 & $4.23^{+0.04}_{-0.04}$ & $0.01^{+0.14}_{-0.16}$ & 0, 0, 0 \\ \\
    4916039 & $0.49881^{+0.00012}_{-0.00013}$ & $0.01043^{+0.00004}_{-0.00004}$ & $101^{+27}_{-24}$ & $32.4^{+2.7}_{-2.7}$ & $4809^{+97}_{-91}$ & $0.61^{+0.04}_{-0.04}$ & 15.44 & $4.59^{+0.03}_{-0.03}$ & $-0.05^{+0.13}_{-0.13}$ & 0, 0, 0 \\ \\
    4930913 & $0.80635^{+0.00019}_{-0.00022}$ & $0.01638^{+0.00006}_{-0.00007}$ & $95^{+134}_{-53}$ & $16.4^{+1.9}_{-1.9}$ & $5726^{+77}_{77}$ & $0.90^{+0.07}_{0.05}$ & 13.96 & $4.35^{+0.13}_{0.11}$ & $-0.12^{+0.15}_{0.15}$ & 0, 0, 0 \\ \\
    4945547 & $0.56461^{+0.00016}_{-0.00015}$ & $0.01217^{+0.00005}_{-0.00005}$ & $66^{+22}_{-17}$ & $24.9^{+2.0}_{-2.0}$ & $4954^{+98}_{-95}$ & $0.75^{+0.03}_{-0.04}$ & 14.46 & $4.59^{+0.02}_{-0.03}$ & $-0.16^{+0.12}_{-0.13}$ & 0, 0, 0 \\ \\
    4996173 & $0.47867^{+0.00008}_{-0.00008}$ & $0.01386^{+0.00004}_{-0.00004}$ & $43^{+12}_{-9}$ & $9.1^{+0.6}_{-0.6}$ & $6712^{+142}_{-140}$ & $1.55^{+0.07}_{-0.07}$ & 10.53 & $4.04^{+0.03}_{-0.03}$ & $0.17^{+0.13}_{-0.16}$ & 0, 0, 0 \\ \\
    \vdots & \vdots & \vdots & \vdots & \vdots & \vdots & \vdots & \vdots & \vdots & \vdots & \vdots \\ \\
    \hline
    \end{tabular}
    \label{tab:all_sample}
\end{table}

We present in Table~\ref{tab:all_sample} the overall properties of 16 out of our 88 selected candidates likely to host non-transiting companions. The orbital period and amplitude are derived from this work. The effective temperature $T_{\rm eff}$, surface gravity log \textit{g}, metallicity and mass measurements parameters are primarily sourced from \citet{Berger2020} while, for stars not included in this catalogue, the properties are taken from \citet{Mathur2017}. The rotation periods are taken from \citet{Santos2019, Santos2021} and the \kep apparent magnitudes Kpmag come from \citet{2011Brown}. The semi-major axis is calculated with the Kepler's third law, assuming the planetary mass is negligible compared to the stellar mass.

The flag column contains a sequence of three binary values (0 or 1) for each entry. Each digit represents a separate flag. If the first value is set to 1, it indicates that tidal forces outweighs atmospheric processes and Doppler boosting in the orbital signal. The second value, if set to 1, means that the target's orbital period is a harmonic of the stellar rotation period (see Appendix~\ref{app : 2ndharmonic systems}). Finally, the third value indicates whether an exoplanet has already been found orbiting around the targeted star, but at a different orbital period than the one detected in this work (see Appendix~\ref{appendix:appendixB}). In such cases, the value is set to 1. The most common set of flags is 0,0,0, indicating that the majority of the sample does not exhibit these characteristics.  

We estimate the uncertainties on the orbital period $P_{\rm orb}$ and the periodogram amplitude $A_{\rm periodogram}$ of the modulation by modelling the main peak attributed in the PSD to the non-transiting planet with a Gaussian profile with central frequency $\nu_0 = 1 / P_{\rm orb}$, height $H$, and standard deviation $\sigma$. Using the \texttt{pyMC} software \citep{pymc}, we perform a Hamiltonian Monte-Carlo exploration of the distribution of the set $\theta = \{P_{\rm orb}, H, \sigma \}$, assuming a $\chi^2$ with two-degrees of freedom as likelihood function for the PSD \citep{Woodard1984}. The periodogram amplitude $A_{\rm periodogram}$ of the modulation, in ppm, is computed as $A_{\rm periodogram}^2 = 2 \sqrt{\pi} \sigma H$. The parameter estimate is taken as the median of the distribution. The lower uncertainty is computed as the difference between the median value of the distribution and the 16th percentile, while the upper uncertainty is computed as the difference between the median value of the distribution and the 84th percentile.

\twocolumn
\section{Systems with an orbital period corresponding to a harmonic of the stellar rotation period \label{app : 2ndharmonic systems}}

In our 88 stars catalogue, the orbital period of 3 candidates corresponds to a harmonic of the stellar surface rotation period.

\begin{figure}[htbp]
   \includegraphics[width=0.4\textwidth]{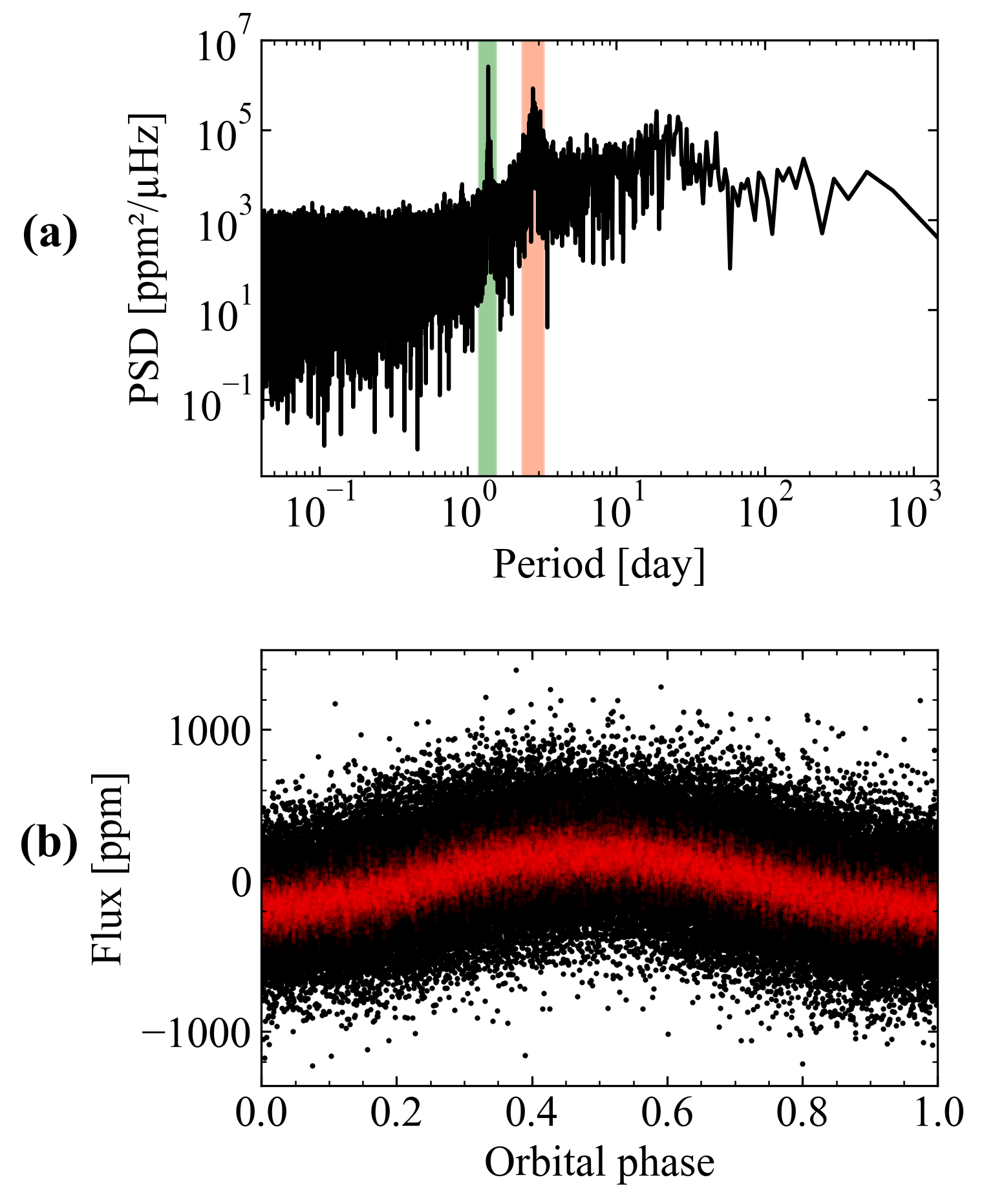}
   \caption{(a) PSD of KIC 7801559. The \prot, shown in the orange area, was measured by \citet{Santos2021}. The green area shows the high-amplitude signal peak at around 1.34 days. (b) Phase-folding analysis on the light curve of KIC 7801559. The red curve is a smoothing of the black curve, obtained with a triangular filter.}
   \label{fig:KIC_7801559}
\end{figure}

\renewcommand{\arraystretch}{0.7}
\begin{table}[htbp]\small
    \centering
    \caption{Candidates with an orbital period corresponding to a harmonic of the stellar rotation period}
    \begin{tabular}{c c c c}
    \hline \hline \\
    KIC & $P_{\rm orb}$ [days]& $P_{\rm rot}$ [days] & Harmonic\\ \\
    \hline \\
    5560359 & $1.96325^{+0.00081}_{-0.00073}$ & $3.9^{+0.5}_{-0.5}$ & 2\\ \\
    7801559 & $1.37431^{+0.00076}_{-0.00051}$ & $2.8^{+0.2}_{-0.2}$ & 2\\ \\
    8193786 & $2.18821^{+0.00120}_{-0.00149}$ & $6.4^{+0.6}_{-0.6}$ & 3\\ \\
    \hline \\
    \end{tabular}
    \tablefoot{The rotation periods are taken from the catalogues by \citet{Santos2019,Santos2021}.}
    \label{tb:2ndharmonic systems}
\end{table}

Panel (a) of Fig.~\ref{fig:KIC_7801559} presents the PSD of KIC 7801559, an F-type star with a \prot of 2.8 days. The companion's orbital period detected in this work corresponds to the second harmonic of the stellar rotation period. On panel (a), the high-amplitude peak corresponding to the orbital signal (in the green area), seems embedded in the rotational harmonic pattern. Nevertheless, the signal shows a very stable periodic behaviour, distinct from the quasi-periodic signature observed in the case of spot modulation, both in the PSD and the light curve phase-folding in panel (b) of Fig.~\ref{fig:KIC_7801559}. 
Therefore, we choose to keep this candidate in our catalogue and suggest that this could be the signature of systems in a spin-orbit resonance. This verification was also done for KIC 5560359 and KIC 8193786. The companion orbital period and the stellar rotation period of these candidates are shown in Table~\ref{tb:2ndharmonic systems}.

\section{Systems with a confirmed exoplanet \label{appendix:appendixB}}

In our 88 stars sample, we detected 2 stars with previously confirmed exoplanets, but at an orbital period different from the one detected in this study. The first star is KIC 10027247, an M-type star with a super-Earth exoplanet, named Kepler-1229 b \citep{2016Morton}. Its mass is 2.54 $\rm M_\oplus$, and it takes 86.8 days to complete one orbit of its star at a distance of 0.3006 Astronomical Unit (AU). Notably, its orbit is within the habitable zone \citep{2017Torres}.
The second star in our catalogue that already hosts a confirmed exoplanet is KIC 9717943, an F-type star. Kepler-1169 b is a terrestrial exoplanet with a mass of 0.779 $\rm M_\oplus$ \citep{2016Morton}. Its orbital period is 6.11 days, and it is located at 0.0709 AU from its host star. 
The properties of these two systems are shown in Table~\ref{tb:2systemesdiscoverd}.

\renewcommand{\arraystretch}{0.7}
\begin{table}[htbp]\small
    \centering
    \caption{Stellar parameters of two candidates with an already confirmed exoplanet.}
    \begin{tabular}{c  c  c }
    \hline \hline \\
    KIC & 9717943 & 10027247 \\ \\ \hline \\
    $T_\text{eff}$ [K] & $6000^{+106}_{-105}$ & $3576^{+75}_{-73}$ \\ \\
    Stellar mass [$\rm M_\odot$] & $1.17^{+0.06}_{-0.07}$ & $0.48^{+0.01}_{-0.01}$ \\ \\
    Kpmag [mag] & 12.72 & 15.47 \\ \\
    $\log\,g$ [dex] & $4.34^{+0.03}_{-0.03}$ & $4.67^{+0.01}_{-0.01}$ \\ \\
    $\rm [Fe/H]$ [dex] & $0.21^{+0.14}_{-0.15}$ & $0.21^{+0.12}_{-0.12}$ \\ \\
    Amplitude [ppm] & $35^{+16}_{-11}$ & $126^{+25}_{-18}$ \\ \\
    $P_\text{rot}$ [days] & $12.5^{+1.6}_{-1.6}$ & $17.5^{+1.5}_{-1.5}$ \\ \\
    semi-major axis [AU] & $0.03092^{+0.00020}_{-0.00020}$ &  $0.01078^{+0.00004}_{-0.00004}$ \\ \\
    $P_\text{orb}$ [days] & $1.83809^{+0.00097}_{-0.00097}$ & $0.58875^{+0.00013}_{-0.00013}$ \\ \\
    \hline \\
    Planet name & Kepler-1169 b & Kepler-1229 b \\ \\
    Confirmed planet's $P_\text{orb}$ [days]& $6.11009^{+0.00003}_{-0.00003}$ & $86.82899^{+0.00107}_{-0.00107}$ \\ \\
    \hline \\
    \end{tabular}
    \tablefoot{The $T_{\rm eff}$, $\textrm{log} \ g$, metallicity, and mass measurements are taken from \citet{Mathur2017}. The rotation periods and \kep apparent magnitudes are taken from \citet{Santos2019,Santos2021} while the orbital periods of the confirmed planets are from \citet{2016Morton}. The semi-major axis is calculated with the Kepler's third law, assuming the planetary mass is negligible compared to the stellar mass. The $P_{\rm orb}$ value was determined in this work for a newly identified, non-transiting object.}
    \label{tb:2systemesdiscoverd}
\end{table}

\end{appendix}
\end{document}